\documentclass{aa}  

\usepackage{color}

\usepackage{hyperref}
\usepackage{graphicx}
\usepackage{txfonts}
\begin{document}

   \title{Cepheid Metallicity in the Leavitt Law (C--MetaLL) survey: VIII. High-Resolution IGRINS Spectroscopy of 23 Classical Cepheids: Validating NIR Abundances.}
\titlerunning{C-MetaLL Survey VIII}
   \subtitle{}

   \author{G. Catanzaro\inst{1}
          \and
          A. Bhardwaj\inst{2}
          \and
          V. Ripepi\inst{3}
          \and
          E. Trentin\inst{3}
          \and
          M. Marconi\inst{3}
          \and
          M. Romaniello\inst{4}
          \and
          N. Matsunaga\inst{7,8}
          \and
          G. De Somma\inst{3,5}
          \and
          T. Sicignano\inst{3,4,5,6}
          \and
          I. Musella\inst{3}
          \and
          Y. Soung-Chul\inst{9}
           }

   \institute{INAF-Osservatorio Astrofisico di Catania, Via S.Sofia 78, 95123, Catania, Italy  \\
              \email{giovanni.catanzaro@inaf.it}
         \and 
         Inter-University Center for Astronomy and Astrophysics (IUCAA), Post Bag 4, Ganeshkhind, Pune 411 007, India
         \and
         INAF-Osservatorio Astronomico di Capodimonte, Salita Moiariello 16, 80131, Naples, Italy
         \and
         European Southern Observatory, Karl-Schwarzschild-Strasse 2, 85748 Garching bei München, Germany 
         \and
         Istituto Nazionale di Fisica Nucleare (INFN)-Sez. di Napoli, Via Cinthia, 80126 Napoli, Italy
         \and 
         Scuola Superiore Meridionale, Largo San Marcellino10 I-80138 Napoli, Italy
         \and
         Department of Astronomy, School of Science, The University of Tokyo, 7-3-1, Hongo, Bunkyo-ku, Tokyo 113-0033, Japan
         \and
         Laboratory of Infrared High-Resolution spectroscopy (LiH), Koyama Astronomical Observatory, Kyoto Sangyo University, Motoyama, Kamigamo, Kita-ku, Kyoto 603-8555, Japan
         \and
         Korea Astronomy and Space Science Institute, Daedeokdae-ro 776, Yuseong-gu, Daejeon 34055, Republic of Korea
             }

   \date{Received ; accepted }

 
  \abstract
   {Classical Cepheids are fundamental primary distance indicators and crucial tracers of the young stellar population in the Milky Way and nearby galaxies. While most chemical abundance studies of Cepheids have been carried out in the optical domain, near-infrared (NIR) spectroscopy offers unique advantages in terms of reduced extinction and access to new elemental tracers.}
   {Our goal is to validate NIR abundance determinations against well-established optical results and to explore the diagnostic power of previously unexplored NIR lines. NIR spectroscopy suffers far less from interstellar extinction than optical observations, allowing us to probe Cepheids at larger distances and in heavily obscured regions of the Galaxy. Moreover, the H and K bands provide access to diagnostic lines of elements (e.g. P, K, and Yb) not available in the optical domain. }
   {We acquired high-resolution (R$\approx$45,000) spectra of 21 Galactic and 2 Large Magellanic Clouds Classical Cepheids with the high-resolution Immersion Grating Infrared Spectrometer (IGRINS) in the H- and K-bands. Effective temperatures were derived from both a photometric approach and line-depth ratios, while gravities and microturbulent velocities were estimated using empirical calibrations and statistical constraints. Abundances of 16 elements were determined through full spectral synthesis in LTE. We performed extensive error analysis and compared our results with previous optical studies of the same stars.}
   {We find excellent agreement ($\Delta$[Fe/H]\,$\le$\,0.02 dex, $\sigma \approx$\,0.07 dex) between our NIR abundances and optical literature values, confirming the reliability of IGRINS-based measurements. The derived abundance gradients in the Galactic disk are fully consistent with previous optical determinations, with slopes of –0.06, –0.05, and –0.05 dex kpc$^{-1}$ for Fe, Mg, and Si, respectively. For the first time we provide homogeneous determinations of P, K, and Yb abundances from NIR lines for Classical Cepheids, finding trends consistent with Galactic chemical evolution models. Moreover, the two LMC Cepheids included in our sample, previously analyzed in the optical, provide a direct benchmark confirming the accuracy of NIR abundance determinations in extragalactic, metal-poor environments.}
   {Our study demonstrates that high-resolution NIR spectroscopy of Cepheids yields robust abundances, fully compatible with optical results, and provides access to additional elements of nucleosynthetic interest. These results pave the way for future large-scale NIR surveys of Cepheids with facilities such as MOONS, ELT, and JWST, crucial for tracing the chemical evolution of the Milky Way and nearby galaxies in heavily obscured regions.}


   \maketitle
%
\begin{table*}
\caption{Logbook for our sample of classical Cepheids.}
\label{tab:logbook}
\centering
\begin{tabular}{lcccccrr}
\hline
\hline
   ID             &    $\alpha$ &    $\delta$    &    V      &    H    &     K  &   R$_{GC}$~~ & Periods  \\
                  &             &                &           &         &        &     (Kpc)    &  (days)  \\
\hline        
AQ Car            & 10:21:22.97 & $-$61:04:26.74 &  8.84     &  6.81   &   6.62 &  7.998 &  9.769 \\
AQ Pup            & 07:58:22.09 & $-$29:07:48.34 &  8.54     &  5.33   &   5.09 &  9.711 & 30.159 \\
DR Vel            & 09:31:40.98 & $-$49:39:18.02 &  9.25     &  5.95   &   5.73 &  8.398 & 11.198 \\
RV Sco            & 16:58:19.74 & $-$33:36:32.77 &  6.61     &  4.82   &   4.63 &  7.465 &  6.061 \\
S Nor             & 16:18:51.84 & $-$57:53:59.26 &  6.49     &  4.50   &   4.22 &  7.537 &  9.754 \\
SS Cma            & 07:26:07.19 & $-$25:15:26.44 &  9.84     &  6.96   &   6.72 &  7.572 &  6.323 \\
S Tra             & 16:01:10.71 & $-$63:46:35.53 &  6.41     &  4.78   &   4.59 & 10.385 & 12.352 \\
SV Vel            & 10:44:56.35 & $-$56:17:22.39 &  7.91     &  5.85   &   5.65 &  7.958 & 14.096 \\
SV Vul            & 19:51:30.91 &   +27:27:36.84 &  6.74     &  3.93   &   3.81 &  7.630 & 44.914 \\
T Mon             & 06:25:13.00 &   +07:05:08.56 &  5.98     &  3.69   &   3.58 &  9.496 & 27.030 \\
U Aql             & 19:29:21.36 & $-$07:02:38.66 &  6.61     &  4.12   &   3.98 &  7.711 &  7.023 \\
U Car             & 10:57:48.19 & $-$59:43:55.88 &  6.11     &  3.94   &   3.79 &  7.900 & 38.791 \\
U Sgr             & 18:31:53.33 & $-$19:07:30.26 &  6.68     &  4.13   &   4.00 &  7.660 &  6.745 \\
V0636 Sco         & 17:22:46.48 & $-$45:36:51.40 &  6.74     &  4.69   &   4.47 &  7.460 &  6.797 \\
VZ Pup            & 07:38:35.23 & $-$28:29:58.60 &  10.15    &  6.66   &   6.46 & 10.947 & 23.172 \\
W Gem             & 06:34:57.45 &   +15:19:49.70 &  6.54     &  4.77   &   4.58 &  9.279 &  7.911 \\
WX Pup            & 07:41:59.01 & $-$25:52:34.25 &  9.09     &  6.71   &   6.54 &  9.850 &  8.936 \\
WZ Sgr            & 18:16:59.71 & $-$19:04:32.99 &  7.45     &  5.05   &   4.78 &  6.451 & 21.847 \\
X Pup             & 07:32:47.04 & $-$20:54:34.88 &  8.46     &  5.41   &   5.22 &  9.494 & 25.971 \\
XX Car            & 10:57:09.22 & $-$65:08:05.12 &  9.42     &  6.90   &   6.69 &  7.747 & 15.714 \\
YZ Car            & 10:28:16.84 & $-$59:21:00.68 &  8.24     &  5.95   &   5.79 &  8.025 & 18.172 \\
\hline
OGLE LMC-CEP-0512 & 04:58:10.80 & $-$69:56:58.78 &  13.13   &  11.04  &   10.64 &        & 39.446   \\   
OGLE LMC-CEP-0992 & 05:07:15.98 & $-$68:53:00.55 &  12.30   &  10.32  &   10.22 &        & 52.875   \\   
\hline
\end{tabular}
\tablefoot{Columns meaning: name of the star in the literature (ID), coordinates referred to 2000.0 equinox ($\alpha$ and $\delta$), magnitudes in V, H, and K band, galactocentric radius \citep[R$_{GC}$, see ]{Gaia2023}, and pulsational periods.}
\end{table*}

\section{Introduction}
A detailed understanding of Classical Cepheids (Cepheids hereafter) is crucial in several astrophysical contexts. First, they are the most important primary distance indicators in the extragalactic distance scale owing to the Leavitt Law \citep{Leavitt1912}, which establishes a period-luminosity (PL) relation. Once calibrated using independent geometric distances—e.g., from trigonometric parallaxes, eclipsing binaries, or water masers—these relations form the first step of the cosmic distance ladder. This primary rung supports the calibration of secondary distance indicators, such as Type Ia supernovae, enabling distance measurements to galaxies in the Hubble flow \citep[see e.g.][]{Freedman2011,Riess2016,Riess2021}.

Cepheids are also excellent tracers of young stellar populations. They enable detailed studies of metallicity gradients along the disk and spiral arms of the Milky Way \citep[MW; see e.g.,][ and reference therein]{Andrievsky2002,luck2011b,Pedicelli2009,Genovali2015,Lemasle2018,luck2018,Ripepi2022a,daSilva2023,Trentin2023a,Trentin2024}, as well as investigations into the three-dimensional geometry, age, and kinematics of the Galactic disk and the Magellanic Clouds \citep[e.g.,][]{Chen2019,Skowron2019,Subramanian2015,Ripepi2017,Desomma2021,Poggio2021,Lemasle2022,Ripepi2022b,drimmel2024,Desomma2025}. Moreover, from theoretical studies of their stellar pulsation,  independent constraints on the stellar masses can be derived \citep[see e.g.,][ and reference therein]{marconi2020}.

While most chemical abundance studies of Cepheids have been performed in the optical, near-infrared (NIR) spectroscopy offers distinct advantages. The significantly reduced interstellar extinction in the NIR—e.g., K-band absorption is roughly ten times lower than in the V band \citep{Cardelli1989}—makes it possible to observe Cepheids in heavily obscured regions, such as the inner Galactic disk or dust-embedded star-forming complexes. Furthermore, the NIR spectral window provides access to unique diagnostic lines of elements like phosphorus (P), potassium (K), and ytterbium (Yb), which are challenging to measure in the optical. High-resolution spectrographs like Immersion GRating INfrared Spectrometer (IGRINS), which cover the $H$ and $K$ bands simultaneously, allow precise atmospheric and abundance analyses for Cepheids in regions where optical observations are impractical.

In this work, we present high-resolution IGRINS spectra of 23 classical Cepheids and perform the first homogeneous abundance analysis of P, K and Yb in the NIR for such a sample. 
The targets of this proposal are composed of 21 well-known Galactic Cepheids and two Large Magellanic Cloud (LMC) Cepheids. The former are part of the sample observed photometrically by HST and used in the local calibration (using the $GAIA$ mission parallaxes) of the Period-Wesenheit\footnote{The Wesenheit magnitude is reddening-free by construction, as long as the reddening law is known \citep{Madore1982}.} relation in the F160W, F555W, F814W bands \citep[see][]{Riess2021,Bhardwaj2023}. The two LMC Cepheids were included as a first-step test of the reliability of NIR-based abundance measurements in extragalactic, low-metallicity environments.

Twelve stars form the full sample (reported in Sect.~\ref{sec_results}) have also been observed by us in the optical bands with the 3.6~m Canada-France-Hawaii Telescope \citep[CFHT, see][and below]{Bhardwaj2023} to allow for a direct cross-validation of the NIR-optical results on the same objects. 

In this respect, this work can be regarded as a pilot project aimed at demonstrating the potential of high-resolution NIR spectroscopy of Cepheids, both for distance scale applications and as tracers of young stellar populations. While previous NIR studies of Cepheids have been carried out, such as those by \citet[][R$\approx$3000, covering up to the K band]{inno2019} and \cite[][R$\approx$28000, limited to the J and H bands]{Matsunaga2023}, none has reached the spectral resolution and simultaneous H–K coverage provided by IGRINS (R$\approx$45000). For the first time, high-resolution spectra of Classical Cepheids in the full H and K bands allow for detailed abundance determinations and robust comparisons with optical results. Indeed, in the context of the   
C-MetaLL project\footnote{https://sites.google.com/inaf.it/c-metall/home}  \citep[Cepheid — Metallicity in the Leavitt Law; see][for details]{Ripepi2021a,Trentin2023a,Bhardwaj2024,Trentin2024,Ripepi2025}, we aim at measuring chemical abundances for a large number of Galactic DCEPs using HiRes spectroscopy, and specifically at extending the iron abundance range into the metal-poor regime, particularly [Fe/H]$<-0.4$ dex. 

The most metal-poor Cepheids are located at large galactocentric radii, often in regions where interstellar extinction dims significantly these objects in the optical, while they are still bright in the NIR regime and within reach of instruments such as IGRINS.  

The structure of the paper is the following: in Sect.~\ref{sec_obs} we present our data and observations; while in Sect.~\ref{sec_atmo} and Sect.~\ref{sec_abundances} we describe the data analysis and abundances estimation, respectively. Results of our analysis are presented in Sect.~\ref{sec_results}, demonstrating the power of NIR spectroscopy to complement and extend optical studies in tracing Galactic chemical evolution and discussed in Sect.~\ref{sec_discussion}. We give our conclusion in Sect.~\ref{sec_conclusion}.

\section{Observations and data analysis}
\label{sec_obs}
We present high-resolution NIR spectroscopic observations of 23 classical Cepheids, comprising 21 located in the Milky Way and 2 in the Large Magellanic Cloud.
The targets are listed in Table~\ref{tab:logbook} together with the log of the observations. All data were acquired using the IGRINS instrument \citep[see][for a detailed description]{yuk2010,gully2012,moon2012,park2014,jeong2014} mounted on the Gemini South telescope \footnote{PI: Bhardwaj, GS-2021-Q-218}.

IGRINS is a compact, high-resolution spectrograph that employs a silicon immersion grating for primary dispersion, coupled with individual volume phase holographic (VPH) gratings as cross-dispersers in separate optical arms. This configuration enables simultaneous coverage of the H (1.49–1.80 $\mu$m) and K (1.96–2.46 $\mu$m) bands across a continuous wavelength range of 1.45–2.45 $\mu$m in a single exposure, achieving a spectral resolution of R\,=\,45000. The absence of moving cryogenic components ensures consistent spectral formatting and instrumental stability across all observations \citep{mace2016,mace2018}.

Data reduction was performed using the IGRINS pipeline package \citep[PLP][]{jae_joon_lee_2017_845059}, which includes flat-fielding, sky subtraction, wavelength calibration (via OH airglow lines and Th-Ar arcs), and optimal extraction of 1D spectra. The final signal-to-noise ratios (S/N) range from 80 to 160 per pixel across the sample, assessed in continuum regions free of telluric or stellar absorption features.

Telluric contamination was rigorously mitigated using observations of A0V standard stars obtained at airmasses closely matched ($\pm$0.1) to those of the target Cepheids.

   \begin{figure}
    \centering
    \includegraphics[width=9cm]{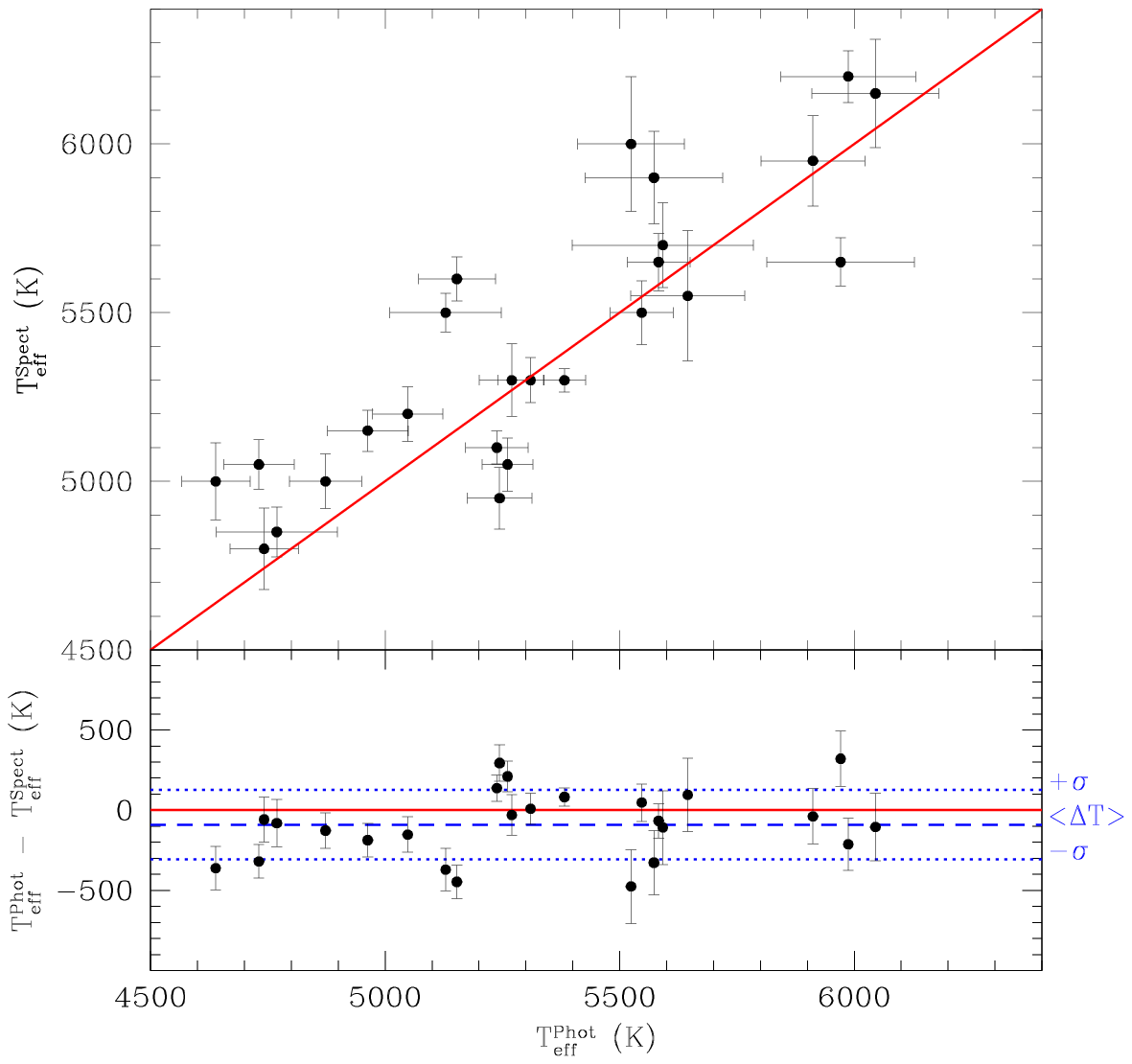}
    \caption{{\it Upper panel:} comparison between effective temperatures derived from spectroscopy and photometry, with the red solid line representing the bisecting line. {\it Bottom panel:} Residuals, computed as $\Delta T_{\text{eff}}$ = $T_{eff}^{Phot}- T_{eff}^{Spec}$, show a mean offset of $-90 \pm 200$~K. With a solid red line we represent null difference ($\Delta T_{\text{eff}} = 0$), while with a blue dashed line the mean residual ($-90$~K). We also show, limited by blue dotted lines, the region of $\pm 1 \sigma$ scatter.}
    \label{fig:compTeff}
   \end{figure}

\section{Atmospheric parameters} \label{sec_atmo}

The effective temperatures ($T_{\rm eff}$) for our sample of classical Cepheids were derived through two independent methodologies: a photometric approach leveraging the stars' intrinsic variability and a spectroscopic technique based on line-depth ratios (LDRs) of carefully selected absorption features.

The photometric method relies on phase-resolved photometry from the Gaia $BP$ and $RP$ bands, calibrated against established colour-temperature relations for Cepheids from \citet{Mucciarelli2021}. $E(B-V)$ values are taken from the literature \citep{Trentin2024}. Thanks to the lightcurves in the $BP$ and $RP$ bands, we estimated the phase-dependent de-reddened colour and calculated the temperature. The spectroscopic method is based on the LDR calibrations of \citet{Kovtyukh2022} and \citet{Afsar23}, which exploit the temperature sensitivity of line pairs (e.g., Ti/Fe blends vs. V I lines) in the $H$- and $K$-bands. This technique provides direct, reddening-insensitive $T_{\rm eff}$ measurements tied to the stellar atmosphere.

Both sets of temperatures are compiled in Table~\ref{tab:param}, with uncertainties of 50$ - $300 K (spectroscopic) and 50$ - $190 K (photometric). As demonstrated in Fig.~\ref{fig:compTeff}, the methods show agreement across the sample, with a mean residual $\Delta T_{\rm eff}=-90\pm 200$ K (photometric minus spectroscopic) and no systematic trends. Considering that the spectroscopic method is less affected by extinction uncertainties, we use the spectroscopically derived $T_{\rm eff}$ values for our subsequent chemical and dynamical analyses. This agreement confirms that line-depth ratios (LDRs) reliably trace Cepheid temperatures in the near-infrared.

\subsection{Limitations and validation of gravity and microturbulence determinations}

The determination of surface gravity ($\log g$) and microturbulent velocity ($\xi$) in our NIR analysis presents specific challenges. In the optical domain, $\log g$ is typically constrained by enforcing ionization equilibrium between Fe I and Fe II lines. However, the NIR spectral window lacks suitable Fe II lines, making this direct method unfeasible. Similarly, a reliable spectroscopic determination of $\xi$ from NIR spectra is compromised by the prevalence of saturated lines, which weakens the sensitivity of the line strength to the microturbulence parameter.

\begin{figure}
    \centering
    \includegraphics[width=9cm]{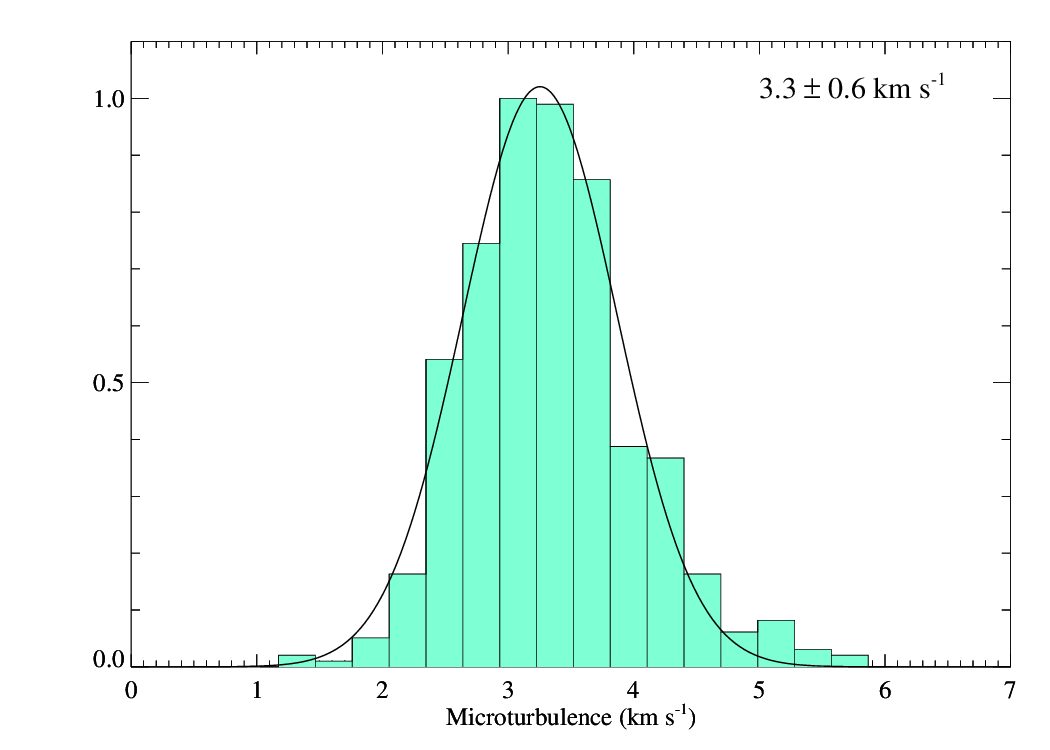}
    \caption{Distribution of microturbulence velocities for the sample of 538 classical Cepheids from the C-MetaLL survey. The Gaussian fit to the distribution yields $\xi$\,=\,3.3\,$\pm$\,0.6 km s$^{-1}$, as labelled.}
    \label{fig:micro}
   \end{figure}

To overcome these limitations, we adopted well-established empirical and statistical approaches. For $\log g$, we used the calibrated relation from \citet{Elgueta2021} and \citet{Matsunaga2023}:
\begin{equation}
\log g = 6.483\log_{10}(T_{\rm eff}/5800) - 0.775\log_{10}P + 2.475
\end{equation}
This relation is based on a large sample of well-characterised Cepheids from \citet{luck2018} and shows minimal dispersion (0.108 dex), providing a robust and consistent estimate. For microturbulence, we employed a statistical constraint derived from the full C-MetaLL sample of 538 spectra of classical Cepheids analyzed so far by us  \citep{Ripepi2021a, Trentin2023a, Trentin2024}. The distribution of $\xi$ measurements was fitted with a Gaussian profile, yielding a characteristic value of $\xi = 3.3 \pm 0.6~{\rm km~s^{-1}}$ (see Fig.~\ref{fig:micro}), which we adopted for all stars in the sample.

We acknowledge that these indirect approaches are a limitation of NIR-based analyses compared to the more direct methods available in the optical. However, the consistency of our final atmospheric parameters and the resulting abundances with extensive optical studies (see Sect.~\ref{sec_results}) validates our methodology. The excellent agreement in the derived abundance gradients and individual element trends demonstrates that these adopted values for $\log g$ and $\xi$ do not introduce significant systematic biases and are appropriate for a homogeneous NIR abundance analysis of Cepheids.

Finally, we determined spectral broadening ($v_{\rm broad}$) through synthetic spectral fitting of multiple Fe I features. Synthetic spectra were convolved with Gaussian profiles of varying widths and compared to observed line shapes. The resulting atmospheric parameters ($T_{\rm eff}$, $\log g$, $v_{\rm broad}$) for all program stars are compiled in Table~\ref{tab:param}.

\begin{table*}
\caption{Derived atmospheric parameters for the sample of classical Cepheids.}
\label{tab:param}
\centering
\begin{tabular}{lcccccc}
\hline
\hline
 ID                &    HJD       &    T$_{eff}^{Phot}$     &    T$_{eff}^{Spec}$    &   $\log g$      &           V$_{rad}$  &   v$_{br}$       \\
                   &  (2400000.+) &   (K)                   &   (K)                  &                 &      (km s$^{-1}$)   &   (km s$^{-1}$)  \\
\hline             
AQ Car             & 59576.853883 & 5547\,$\pm$\,~~67 & 5500\,$\pm$\,100  & 1.5\,$\pm$\,0.1 &  ~~58.3\,$\pm$\,0.2 &  13.0\,$\pm$\,0.7 \\
AQ Pup             & 59488.888079 & 4639\,$\pm$\,~~73 & 5000\,$\pm$\,110  & 0.9\,$\pm$\,0.1 &   121.0\,$\pm$\,0.2 & ~~7.0\,$\pm$\,0.5 \\ 
DR Vel             & 59505.861295 & 5048\,$\pm$\,~~75 & 5200\,$\pm$\,~~80 & 1.3\,$\pm$\,0.1 &  ~~94.8\,$\pm$\,0.2 & ~~8.0\,$\pm$\,1.2 \\
RV Sco             & 59695.690020 & 5238\,$\pm$\,~~67 & 5100\,$\pm$\,~~50 & 1.5\,$\pm$\,0.1 &  ~~53.8\,$\pm$\,0.2 &  12.0\,$\pm$\,0.6 \\
S Nor              & 59656.791525 & 5382\,$\pm$\,~~45 & 5300\,$\pm$\,~~50 & 1.5\,$\pm$\,0.1 &  ~~51.5\,$\pm$\,0.2 &  12.0\,$\pm$\,0.8 \\
SS Cma             & 59482.854633 & 5583\,$\pm$\,~~67 & 5650\,$\pm$\,~~80 & 1.8\,$\pm$\,0.1 &   126.1\,$\pm$\,0.2 &  14.0\,$\pm$\,0.8 \\
S Tra              & 59651.740433 & 5261\,$\pm$\,~~54 & 5050\,$\pm$\,~~80 & 1.2\,$\pm$\,0.1 &  ~~74.3\,$\pm$\,0.2 &  15.0\,$\pm$\,2.0 \\
SV Vel             & 59436.455912 & 4963\,$\pm$\,~~86 & 5150\,$\pm$\,~~60 & 1.2\,$\pm$\,0.1 &   121.4\,$\pm$\,0.3 &  13.0\,$\pm$\,2.2 \\
SV Vul             & 59705.903788 & 4731\,$\pm$\,~~75 & 5050\,$\pm$\,~~75 & 0.8\,$\pm$\,0.1 &  ~~77.4\,$\pm$\,0.3 &  15.0\,$\pm$\,0.6 \\
T Mon              & 59651.514273 & 5592\,$\pm$\,193  & 5700\,$\pm$\,125  & 1.3\,$\pm$\,0.1 &   117.2\,$\pm$\,0.3 &  18.0\,$\pm$\,1.0 \\
T Mon              & 59694.465267 & 4769\,$\pm$\,129  & 4850\,$\pm$\,~~75 & 0.9\,$\pm$\,0.1 &   147.5\,$\pm$\,0.3 &  14.0\,$\pm$\,1.2 \\
U Aql              & 59695.852863 & 5987\,$\pm$\,144  & 6200\,$\pm$\,~~75 & 2.0\,$\pm$\,0.1 &  ~~39.4\,$\pm$\,0.2 &  14.0\,$\pm$\,0.7 \\
U Car              & 59603.788465 & 5524\,$\pm$\,114  & 6000\,$\pm$\,300  & 1.3\,$\pm$\,0.2 &  ~~49.4\,$\pm$\,0.4 &  17.0\,$\pm$\,1.5 \\
U Sgr              & 59695.837884 & 5912\,$\pm$\,111  & 5950\,$\pm$\,135  & 1.9\,$\pm$\,0.1 &  ~~45.0\,$\pm$\,0.2 &  15.0\,$\pm$\,0.9 \\
V0636 Sco          & 59692.696240 & 5244\,$\pm$\,~~69 & 4950\,$\pm$\,~~90 & 1.4\,$\pm$\,0.1 &  ~~97.5\,$\pm$\,0.3 &  16.0\,$\pm$\,0.5 \\
VZ Pup             & 59488.882649 & 4873\,$\pm$\,~~77 & 5000\,$\pm$\,~~80 & 1.0\,$\pm$\,0.1 &   138.2\,$\pm$\,0.2 & ~~7.0\,$\pm$\,0.5 \\
W Gem              & 59641.518548 & 6045\,$\pm$\,135  & 6150\,$\pm$\,160  & 1.9\,$\pm$\,0.2 &  ~~96.1\,$\pm$\,0.2 &  14.0\,$\pm$\,0.4 \\
WX Pup             & 59488.866445 & 5270\,$\pm$\,~~69 & 5300\,$\pm$\,110  & 1.5\,$\pm$\,0.1 &   117.7\,$\pm$\,0.2 & ~~7.0\,$\pm$\,0.5 \\
WZ Sgr             & 59695.870539 & 5971\,$\pm$\,157  & 5650\,$\pm$\,~~70 & 1.4\,$\pm$\,0.1 &  ~~13.6\,$\pm$\,0.3 &  18.0\,$\pm$\,0.8 \\
X Pup              & 59488.859321 & 4742\,$\pm$\,~~73 & 4800\,$\pm$\,120  & 0.8\,$\pm$\,0.2 &   157.0\,$\pm$\,0.4 &  21.0\,$\pm$\,0.6 \\
XX Car             & 59438.455383 & 5153\,$\pm$\,~~82 & 5600\,$\pm$\,~~65 & 1.5\,$\pm$\,0.1 &  ~~85.0\,$\pm$\,0.2 &  14.0\,$\pm$\,1.4 \\
YZ Car             & 59578.864370 & 5573\,$\pm$\,147  & 5900\,$\pm$\,140  & 1.6\,$\pm$\,0.1 &  ~~67.5\,$\pm$\,0.2 &  14.0\,$\pm$\,0.6 \\
YZ Car             & 59582.821720 & 5129\,$\pm$\,119  & 5500\,$\pm$\,~~60 & 1.3\,$\pm$\,0.1 &  ~~70.6\,$\pm$\,0.2 &  10.0\,$\pm$\,0.6 \\
\hline                                         
OGLE LMC-CEP-0512  & 59491.680453 & 5645\,$\pm$\,122  & 5550\,$\pm$\,190  & 1.1\,$\pm$\,0.2 &   334.2\,$\pm$\,0.8 &  30.0\,$\pm$\,2.0 \\
OGLE LMC-CEP-0992  & 59588.610443 & 5310\,$\pm$\,~~70 & 5300\,$\pm$\,~~70 & 0.9\,$\pm$\,0.1 &   355.2\,$\pm$\,0.2 &  11.0\,$\pm$\,0.6 \\
\hline
\end{tabular}
\tablefoot{Columns include the star identifier (ID), Heliocentric Julian Date (HJD), effective temperature and uncertainties from photometry (T$_{eff}^{Phot}$) and spectroscopy (T$_{eff}^{Spec}$), surface gravity ($\log g$), radial velocity (V$_{rad}$), and broadening velocity (v$_{broad}$).}
\end{table*}


\section{Abundances} \label{sec_abundances}

   \begin{figure*}
    \centering
    \includegraphics{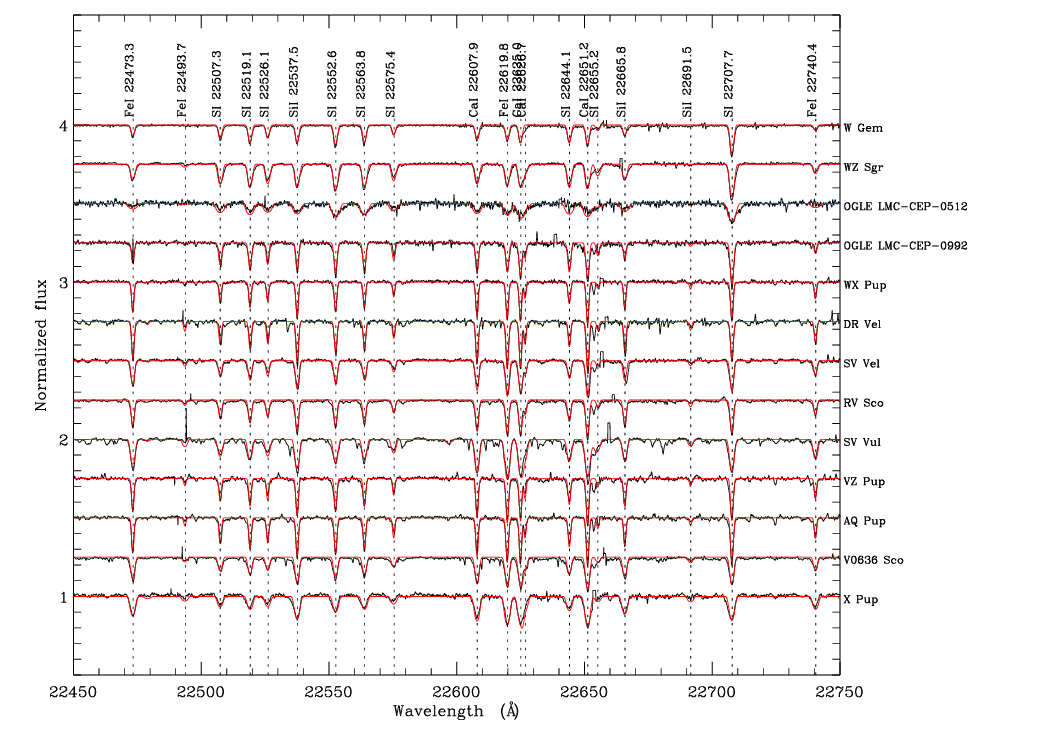}
    \caption{Representative sub-sample of 13 observed spectra (black) in the region between $\lambda$\,= 22450 and 22750 {\AA}. The best-fitting synthetic spectra are overplotted in red. The spectra are ordered, from top to bottom, by decreasing effective temperature (see Table~\ref{tab:param}. The main spectral lines have been identified at the top and highlighted with black dotted vertical lines.}
    \label{fig:spectra}
   \end{figure*}

To mitigate spectral line blending and broadening effects, we employed spectral synthesis techniques. Synthetic spectra were generated through a three-step process:
\begin{enumerate}
    \item \textbf{Atmospheric modeling}: Plane-parallel Local Thermodynamic Equilibrium (LTE) atmospheric structures were computed using \textsc{atlas9} \citep{kurucz1993new}, initialized with stellar parameters from Table~\ref{tab:param};
    \item \textbf{Spectral synthesis}: Synthetic spectra were generated with \textsc{synthe} \citep{kurucz1981solar}. The line lists and atomic parameters adopted originate mainly from \citet{castelli2004spectroscopic}, which provides updated oscillator strengths ($\log gf$) and damping constants for the compilation of \citet{kurucz1995kurucz}. These were supplemented with data from \textsc{vald3} \citep{pisk1995} and cross-verified against the APOGEE DR16 line list \citep{Smith2021}. This multi-source approach ensured completeness for NIR diagnostics and consistency in damping treatments. Finally, the spectral lines selected for this study, together with their atomic parameters, have been reported in Table~\ref{sel_lines};
    
    \item \textbf{Broadening convolution}: Synthetic spectra were convolved with Gaussian kernels to match instrumental resolution ($R \sim 45,000$) and velocity broadening (including rotational and macroturbulent velocities), optimised by fitting unblended metal lines.
\end{enumerate}

Abundances were derived for 16 elements (C, N, Na, Mg, Al, Si, P, S, K, Ca, Ti, Mn, Fe, Ni, Ce, Yb) using a $\chi^2$ minimisation. In particular:
\begin{itemize}
    \item Spectra were divided into 25--50\,\AA\ intervals;
    \item Abundance solutions were obtained by minimizing:
    \[
    \chi^2 = \sum_{i=1}^{n} w_i^2\left[ F_{\text{obs}}(\lambda_i) - F_{\text{synth}}(\lambda_i, A) \right]^2
    \]
    where $A$ is the elemental abundance and $w_i\,=\,1/\sigma_i$ are the weights associated to the $F_{\text{obs}}(\lambda_i)$ (i.e. SNR);
    \item The IDL implementation of the \textsc{amoeba} algorithm \citep{Press1992} performed optimization;
    \item Final abundances represent the median of all valid solutions per element.
\end{itemize}

Uncertainties were estimated via Monte Carlo simulation, perturbing continuum placement ($\pm 0.5\%$) and atmospheric parameters within errors ($T_{\text{eff}} \pm 150$~K, $\log g \pm 0.15$~dex).The $1\sigma$ dispersion of the resulting abundance distributions defines our reported errors. These simulations indicate a typical uncertainty contribution of $\pm$\,0.1 dex. Then, total errors were evaluated by summing in quadrature the value obtained by the error propagation and the standard deviations obtained from the average abundances. For elements with only one measurable spectral line (e.g., phosphorus), we could not estimate the dispersion from multiple lines. In these cases, we conservatively adopted the maximum uncertainty derived from the error propagation on the stellar parameters, which amounts to $\approx 0.15$\,dex. The final list of LTE abundances for all stars is reported in Table~\ref{abund}. All abundances are referred to the solar value \citep{grevesse2011chemical}.

Our abundance analysis is based on the assumption of LTE. While non-LTE effects can be important for certain elements and stellar types, we justify this choice based on the following considerations. The Cepheids in our sample have effective temperatures in the range T$_{\text{eff}} \approx$~4800 -- 6200~K and metallicities $-$0.4$\le$[Fe/H]$\le$0.3 dex, placing them within the parameter space covered by the non-LTE departure coefficient grids of \citet{amarasi2020}. According to their study, non-LTE corrections for many elements in this temperature and metallicity regime are small ($\le$\,0.05~dex) for giants ($\log g \le$\,3.5), especially when differential abundances relative to the Sun are considered. For example, elements such as Si, Ca, Ti, and Fe show negligible non-LTE corrections under these conditions. Although some species like O, Na, and K may exhibit larger departures, the overall impact on our abundance trends and radial gradients is expected to be minor, as confirmed by the good agreement between our near-infrared IGRINS abundances and optical non-LTE literature values (see Sect.~\ref{sec_results}). Therefore, the use of LTE in this work is well justified for the purpose of deriving relative chemical abundances and gradients across the Galactic disc. In Fig.~\ref{fig:spectra} we show an example of our observed (in black) and fitted synthetic (in red) spectra, highlighting some of the identified lines in the range between $\lambda$\,22450 {\AA} and $\lambda$\,22750 {\AA}. 


\section{Results}\label{sec_results}

We present here the chemical abundance results derived from the high-resolution IGRINS spectra of 23 Classical Cepheids. The analysis focuses on elements commonly studied in the optical domain (e.g., Fe, Mg, Si), enabling direct comparisons, as well as on species accessible primarily in the NIR (e.g., P, K, Yb). 

\begin{figure*}
    \centering
    \includegraphics{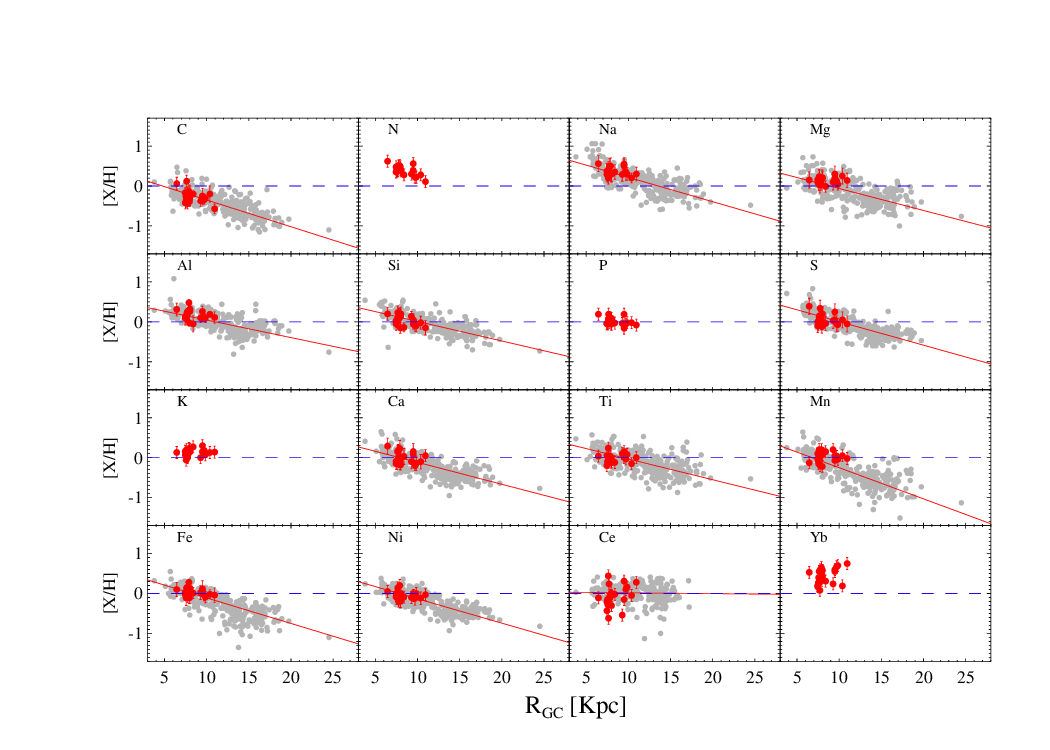}
    \caption{Radial abundance gradients of selected elements in the Galactic disc. Red points represent the galactic cepheids abundances derived in this work from IGRINS H–K spectra, while grey points correspond to optical measurements from the C‑MetaLL sample. Solid red lines indicate the linear fits adopted from \citet{Trentin2024}.}
    \label{gradient}
   \end{figure*}

Fig.~\ref{gradient} shows the radial Galactocentric gradients. The red points correspond to the measurements of galactic cepheids presented in this work, while the grey points represent the C-MetaLL sample, which is not available for all elements. For each element, the linear fits shown in the figure were adopted from \citet{Trentin2024} and are included here for reference. The new measurements broadly follow the trends established by previous optical studies, confirming the presence of negative abundance gradients across the Galactic disc. In particular, the slopes derived in \citet{Trentin2024} for [Fe/H], [Mg/H], and [Si/H] are –0.064, –0.055 and –0.049 dex kpc$^{-1}$, respectively, and our data are fully consistent with these values. Although the radial coverage is comparable to that of previous studies, the present analysis provides an independent validation of these trends based on near-infrared H- and K-band spectroscopy with IGRINS. This supports the reliability of abundance determinations across different spectral ranges.

   \begin{figure*}
    \centering
    \includegraphics{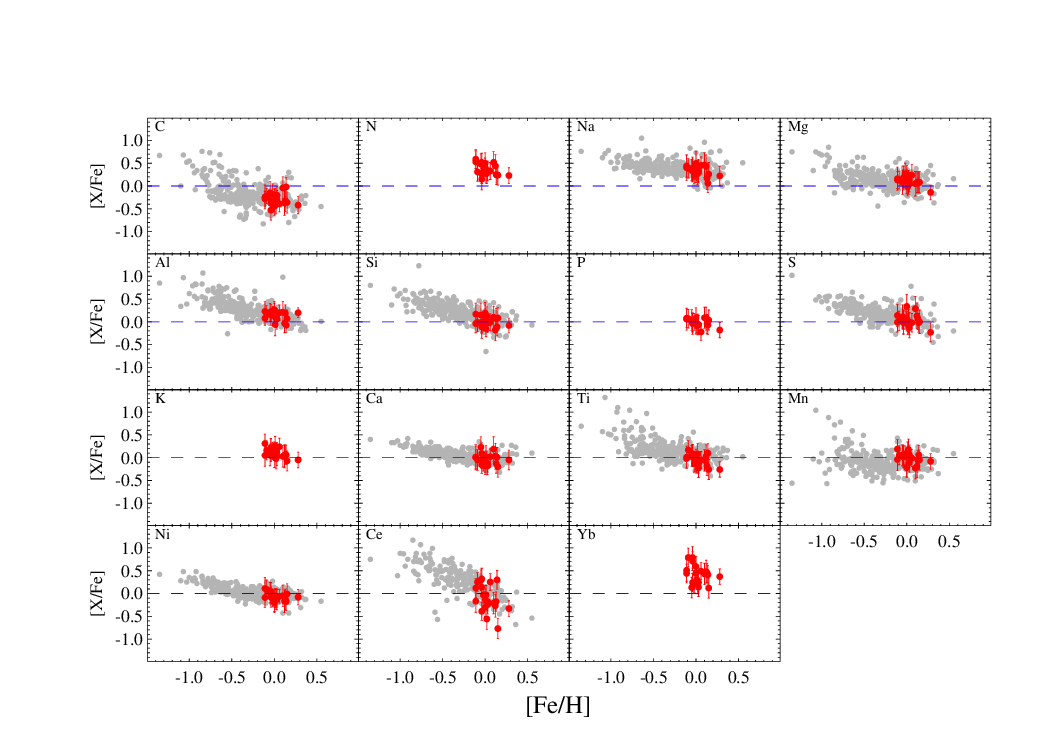}
    \caption{Chemical abundances of elements (in their [X/Fe] form) plotted against iron abundance [Fe/H]. The horizontal dashed line at [X/Fe] = 0 indicates the solar abundance ratio. Meaning of the colors is as in Fig.~\ref{gradient}.}
    \label{gradientFe}
   \end{figure*}
The chemical abundance patterns of our stellar sample are also presented in Fig.~\ref{gradientFe}, where we plot [X/Fe] ratios against metallicity [Fe/H]. The grey points, taken from the C-MetaLL sample, provide a valuable baseline for comparison and establish the general chemical evolution trends in the stellar population.

Our analysis reveals several key features. The overall trend shows the characteristic decline in [X/Fe] with increasing [Fe/H], consistent with the established scenario of Galactic chemical evolution where Type Ia supernovae begin to contribute significantly to iron production after the initial enrichment by core-collapse supernovae (SNeII). The red points, representing our specific measurements, show excellent agreement with the C-MetaLL trends, validating our analysis methodology.

Notably, the spread in [X/Fe] ratios at low metallicities in both our data and the C-MetaLL sample reflects the inhomogeneous nature of chemical enrichment in the early Galaxy. The consistency between our results and the optical measurements strengthens the interpretation that these patterns represent genuine features of Galactic chemical evolution rather than analysis artifacts.

\begin{figure*}
    \centering
    \includegraphics{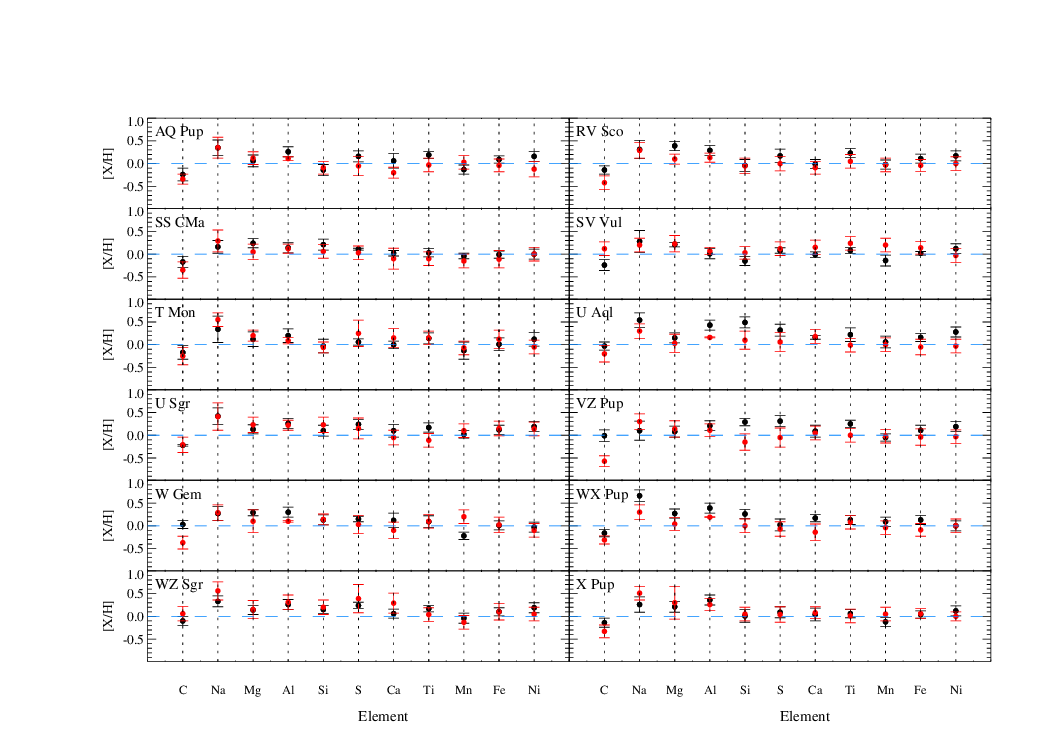}
    \caption{Differences between elemental species (in the x axis) derived in this work from near-infrared IGRINS spectra (black dots) and those published by T24 using optical high-resolution spectra obtained with ESPaDOnS@CFHT (red dots). In the y axis the difference is expressed in the [X/H] form. Each panel corresponds to a star in common between the two samples. }
    \label{comp}
   \end{figure*}

To further assess the reliability of our abundance determinations based on NIR spectra, we compared our results with independent measurements available in the literature for the same stars. Fig.~\ref{comp} presents a direct comparison for 12 stars between the abundances derived in this work and those published in \citet[][Fe only]{Bhardwaj2023} and in \citet[][11 chemical species]{Trentin2024}, based on optical high-resolution spectra obtained with the Echelle SpectroPolarimetric Device for the Observation of Stars (ESPaDOnS\footnote{ https://www.cfht.hawaii.edu/Instruments/Spectroscopy/Espadons/}), mounted at CFHT. ESPaDOnS operates with a spectral resolution of R\,=\,81000 and covers the wavelength range from 3700 to 10500 {\AA}.

The comparison reveals an excellent overall agreement between the two datasets, with a mean absolute scatter of $\approx$0.07 dex, consistent with the typical measurement uncertainties. For iron, we find a negligible median offset of less than 0.02 dex, and the dispersion remains within 0.06 dex. Among the $\alpha$-elements, magnesium, silicon, and calcium also show good consistency, with mean differences of ~0.05 dex or less and no significant systematic trends. These results confirm that the abundances derived from IGRINS H- and K-band spectra are fully compatible with those obtained from optical spectroscopy, thereby validating the reliability of our near-infrared line selection, model atmospheres, and analysis procedures. This agreement is particularly encouraging for future large-scale abundance studies in the NIR domain, especially where optical data may be inaccessible.

   \begin{figure*}
    \centering
    \includegraphics{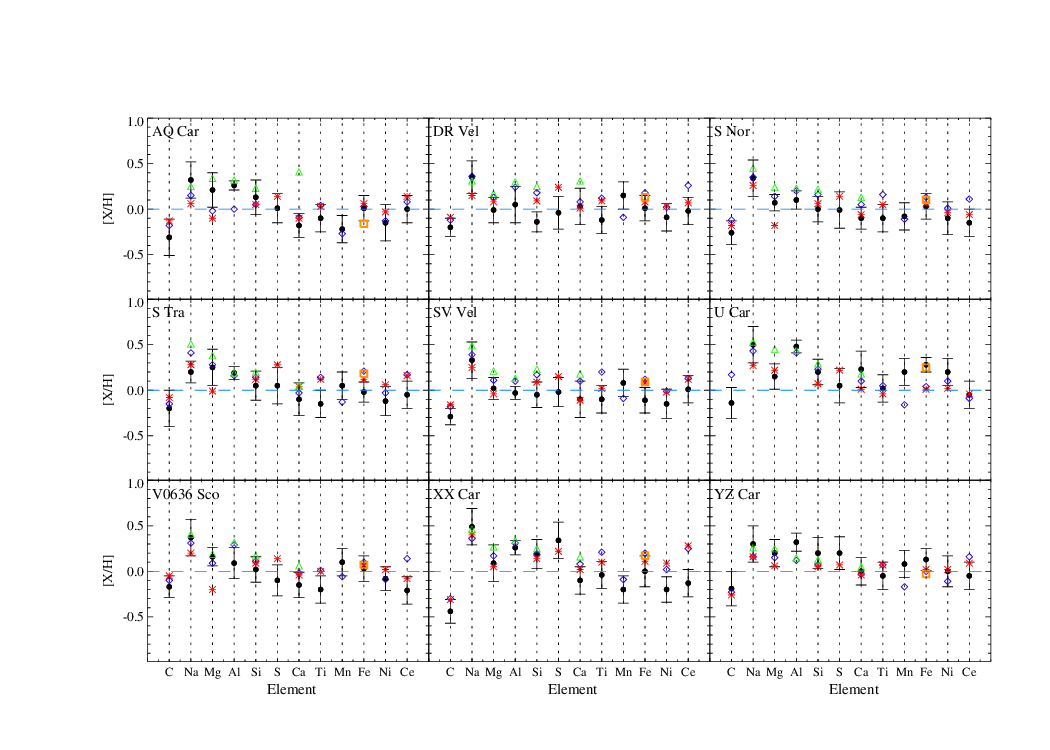}
    \caption{Comparison between elemental abundances derived in this work from IGRINS near-infrared spectra (black circles) and values available in the literature for the same stars. Red asterisks correspond to \citet{luck2011a}, blue diamonds to \citet{luck2011b}, orange squares to \citet{Genovali2014} and green triangles to \citet{Genovali2015}. The dashed blue horizontal line marks the null difference.}
    \label{complit}
   \end{figure*}

   \begin{figure}
    \centering
    \includegraphics[trim=18 200 0 0,width=9cm]{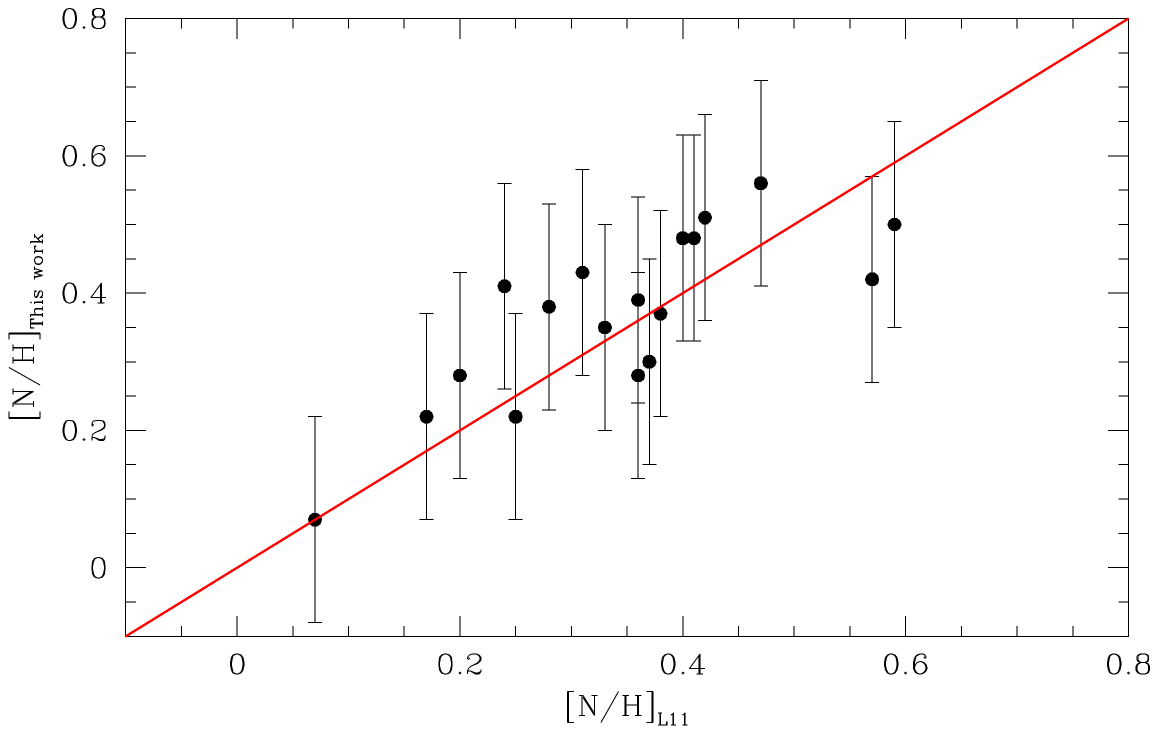}
    \caption{Comparison between nitrogen abundances derived in this work from IGRINS near-infrared spectra (y axis) with those published by \citet{luck2011a} (x axis) from optical spectroscopy, for 18 stars in common. The solid red line indicates the one-to-one relation.}
    \label{compN}
   \end{figure}

We complemented the comparison shown in Fig.~\ref{comp} by comparing our abundance determinations with those available in the literature outside the C-MetaLL project. We retrieved abundances from different sources, which implement independent analyses such as  \citet{luck2011a,luck2011b,Genovali2014,Genovali2015}. Fig.~\ref{complit} shows this comparison. These literature studies differ in terms of spectroscopic resolution, wavelength coverage, line selection, and analysis techniques. Despite this methodological diversity, our results show 1$\sigma$ agreement with the published values. Most data points lie close to the null difference line, with no significant systematic offsets, confirming the reliability of our IGRINS-based near-infrared abundance measurements. This agreement, even in a non-homogeneous comparison, reinforces the robustness of our results and their compatibility with optical literature data.

In Fig.~\ref{compN} we compare the nitrogen abundances derived in this work with near-infrared IGRINS spectra with those published by \citet{luck2011a} for a subset of 18 stars in common. Despite the differences in wavelength domain, instrumentation, and line selection (our abundances being based on the spectral line N\,I 15852.287 {\AA} and those of \citet{luck2011a} derived from optical atomic lines), the agreement between the two datasets is remarkably good. The median difference is only +0.03 dex, with a standard deviation of 0.07 dex, well within the expected combined uncertainties. Most points lie close to the one-to-one relation, and no significant trend with abundance level is observed. This level of consistency confirms that reliable nitrogen abundances can be obtained from near-infrared spectra and further supports the robustness of the methodology adopted in this work.

\section{Discussion}\label{sec_discussion}
Our study demonstrates that near-infrared spectroscopy with IGRINS provides reliable chemical abundances for both well-studied elements and species that are largely inaccessible in the optical domain. The derived $\alpha$-element gradients (Mg, Si, Ca, Ti) confirm the well-established negative slopes across the Galactic disk, in excellent agreement with optical studies. These gradients are a fundamental prediction of inside-out galaxy formation models \citep[e.g.][]{spitoni2019, Grisoni18}, where the inner disk forms stars more rapidly and efficiently, leading to earlier and more pronounced enrichment by SNeII. 
The consistency of our NIR-based gradients with those from optical studies and with chemo-dynamical models provides an independent validation of these theoretical frameworks. It also underscores that NIR spectroscopy is a powerful and reliable tool for probing Galactic chemical evolution, even in regions where optical data are limited by extinction.

In addition to the $\alpha$-elements, our IGRINS spectra enable the study of potassium, which occupies an intermediate position between the light $\alpha$-elements and the heavier neutron-capture species. The [K/H] abundances derived from the NIR lines exhibit a mild correlation with [Fe/H], consistent with the behavior expected for an element produced mainly in massive stars and released into the interstellar medium by SNeII. The lack of significant deviations from the solar-scaled trend suggests that Cepheids accurately trace the present-day potassium content of the Galactic thin disk. This behavior supports theoretical predictions that K nucleosynthesis is sensitive to both progenitor mass and explosion energy, with possible contributions from explosive burning in oxygen- and neon-rich layers \citep[e.g.][]{kobayashi11}. Our results identify potassium as a valuable diagnostic element that bridges the information provided by $\alpha$-elements and heavier r- and s-process tracers such as Yb.

A key advancement of this work is the first homogeneous determination of phosphorus and ytterbium abundances in Cepheids from NIR lines. Our phosphorus measurements, obtained from the \ion{P}{I} line at 16482.92 {\AA} in the H band, are consistent with the methodology of recent high-resolution NIR studies. The work on K giants by \citet{Nandakumar22}, which also used IGRINS, revealed a clear decreasing trend of [P/Fe] with increasing [Fe/H] for disk stars. This behavior, similar to that of $\alpha$-elements like Mg and Si, supports a primary nucleosynthetic origin in SNeII, likely associated with the $\alpha$-rich freeze-out process \citep{Timmes95, Cescutti12}. Recent calibrations of YJ-band P lines in Cepheids \citep{Elgueta24} further highlight the potential of NIR diagnostics for this element. The successful measurement of phosphorus in our Cepheid sample, despite challenges such as telluric contamination and line blending, validates the use of this infrared feature for reliable abundance determinations in luminous, pulsating stars. 

Ytterbium exhibits a more complex behavior. The derived [Yb/Fe] ratios indicate a dominant r-process contribution ($\approx$\,60$\%$), as suggested by their similarity to the [Eu/Fe] trend \citep{Montelius22}. However, the relatively high [Yb/H] values observed in some disk stars imply a non-negligible s-process component from asymptotic giant branch (AGB) stars. This dual origin makes Yb a powerful diagnostic for disentangling the relative contributions of rapid (neutron-star mergers or core-collapse SNe) and slow (AGB) neutron-capture processes to the chemical enrichment of the young Galactic disk. The agreement between our Cepheid Yb abundances and those measured in K giants \citep{Montelius22} confirms that NIR lines can reliably trace this complex element across different stellar populations.

By combining the classical $\alpha$-elements, which trace the yields of core-collapse supernovae, with P (a specific SNeII product), K (a transitional element with explosive nucleosynthetic origin), and Yb (a mixed r-/s-process tracer), our NIR study provides a holistic view of nucleosynthesis in the Galactic disk. The fact that Cepheids, representatives of the young stellar population, show chemical patterns consistent with those of older K giants suggests that the main production channels for these elements have remained relatively stable throughout the recent history of the disk. However, subtle differences in the [Yb/Fe] trend at high metallicity may offer future constraints on the delay-time distribution of r-process events.

Finally, the IGRINS observations of the two Cepheids in the LMC, OGLE LMC-CEP-0512 and OGLE LMC-CEP-0992, represent the first high-resolution NIR spectroscopic analysis of these fundamental variables in an external galaxy. These two stars were also analyzed in the optical by  \citep[][]{Romaniello2022}, allowing for a direct and robust comparison between NIR and optical abundance determinations. The [Fe/H] values we derive ($-$0.39 and $-$0.31 dex, respectively) are fully consistent with their optical metallicities, providing strong validation of the reliability of NIR-based abundance measurements in low-metallicity, extragalactic environments.

\section{Conclusions}\label{sec_conclusion}

We presented a detailed chemical analysis of 23 classical Cepheids observed in the H and K bands with the high-resolution spectrograph IGRINS. From these data, we derived atmospheric parameters and abundances for sixteen chemical species. The effective temperatures obtained from near-infrared line-depth ratios show an excellent agreement with photometric estimates, confirming the reliability of this method for pulsating variable stars.

The abundances derived from the near-infrared spectra are fully consistent with those obtained from optical high-resolution studies, with negligible offsets and dispersions within the observational uncertainties. The $\alpha$-elements (Mg, Si, Ca, and Ti) display negative radial gradients across the Galactic disk that are consistent with previous optical results and with chemo-dynamical models predicting an inside-out formation of the Milky Way. This agreement validates the use of near-infrared spectroscopy as a robust tool for tracing the chemical structure of the Galaxy, even in regions where optical observations are hampered by extinction.

A major outcome of this work is the first homogeneous determination of phosphorus, potassium, and ytterbium abundances in Cepheids from near-infrared spectra. Phosphorus and potassium show behaviours compatible with production in massive stars and SNeII, while ytterbium traces a combination of r- and s-process enrichment, as expected for neutron-capture elements in the Galactic disk. The ability to measure these elements, which are poorly accessible in the optical range, highlights the unique diagnostic potential of high-resolution NIR spectroscopy.

Overall, our analysis demonstrates that IGRINS is a powerful and reliable instrument for abundance studies in Cepheids. Near-infrared spectroscopy not only confirms large-scale chemical trends previously established in the optical but also extends such studies to additional nucleosynthetic tracers and to highly obscured regions of the Milky Way and nearby galaxies. This capability paves the way for future large-scale NIR surveys with facilities such as MOONS, ELT, and JWST, which will be essential to map the chemo-dynamical evolution of our Galaxy fully.

\begin{acknowledgements}
We acknowledge the financial support from INAF (Large Grant MOVIE, PI: Marconi). 
We acknowledge funding from INAF GO-GTO grant 2023 “C-MetaLL - Cepheid metallicity in the Leavitt law” (P.I. V. Ripepi) and Project PRIN MUR 2022 (code 2022ARWP9C) “Early Formation and Evolution of Bulge and HalO (EFEBHO)", PI: Marconi, M.,  funded by European Union – Next Generation EU. A.B. thanks the funding from the Anusandhan National Research Foundation (ANRF) under the Prime Minister Early Career Research Grant scheme (ANRF/ECRG/2024/000675/PMS). G.D.S. acknowledges support from the INAF–ASTROFIT fellowship, from Gaia DPAC through INAF and ASI (PI: M. G. Lattanzi), and from INFN (Naples Section) through the QGSKY and Moonlight2 initiatives.

This research has made use of the
SIMBAD database operated at CDS, Strasbourg, France.
 This work has made use of data from the European Space
Agency (ESA) mission Gaia (https://www.cosmos.esa.int/gaia),
processed by the Gaia Data Processing and Analysis Consortium (DPAC,
https://www.cosmos.esa.int/web/gaia/dpac/consortium). Funding
for the DPAC has been provided by national institutions, in particular, the
institutions participating in the Gaia Multilateral Agreement.
This research was supported by the Munich Institute for Astro-, Particle and BioPhysics (MIAPbP), which is funded by the Deutsche Forschungsgemeinschaft (DFG, German Research Foundation) under Germany´s Excellence Strategy – EXC-2094 – 390783311.
This research was supported by the International Space Science Institute (ISSI) in Bern/Beijing through ISSI/ISSI-BJ International Team project ID \#24-603 - “EXPANDING Universe” (EXploiting Precision AstroNomical Distance INdicators in the Gaia Universe).
\end{acknowledgements}

%
   \bibliographystyle{aa} 
   \bibliography{mybib} 
%

\begin{appendix}

\section{Spectral line list and abundances}    

   \begin{table*}
      \caption[]{List of spectral lines used in this study. For each transition we report the laboratory wavelength, $\log (gf)$, energy (E, in eV) and total angular momentum (J) for lower and upper levels, and radiative, Stark, and Van der Waals damping constant in logaritmic form ( $\log(\gamma_{r})$,$\log(\gamma_{S})$ and $\log(\gamma_{W})$. Blended lines are marked with an asterisk.}
         \label{sel_lines}
\centering
\begin{tabular}{lrrcrccrr}
\hline
\hline
Wavelength  & $\log(gf)$ & E$_i$~  & J$_i$  &  E$_f$~    & J$_f$  & $\log(\gamma_{r})$ & $\log(\gamma_{S})$ &  $\log(\gamma_{W})$ \\
~~~~~~~~({\AA}) & & (eV) & & (eV) & & & & \\
\hline
N\,I    &        &       &     &        &     &      &       &       \\
15852.287   & -0.020 &  12.12 & 1.5 &  12.92 & 1.5 & 0.00 &  0.00 &  0.00 \\
\hline
C\,I    &        &       &     &        &     &      &       &       \\
15852.580   & -0.258 &  9.63 & 2.0 &  10.41 & 3.0 & 0.00 &  0.00 &  0.00 \\
16004.872   &  0.234 &  9.63 & 2.0 &  10.41 & 3.0 & 7.96 & -4.44 & -6.99 \\
16172.819   & -0.988 & 10.40 & 1.0 &  11.17 & 0.0 & 7.95 & -2.66 & -6.62 \\
16415.693   & -1.241 &  9.33 & 1.0 &  10.09 & 2.0 & 8.66 & -4.93 & -7.00 \\
16465.026   & -1.450 &  9.33 & 1.0 &  10.08 & 1.0 & 8.66 & -4.93 & -7.00 \\
16468.533   & -1.156 &  9.33 & 2.0 &  10.08 & 1.0 & 8.66 & -4.93 & -7.00 \\
16470.912   & -1.342 &  9.33 & 0.0 &  10.08 & 1.0 & 8.66 & -4.93 & -7.00 \\
16505.146   & -1.292 &  9.33 & 1.0 &  10.08 & 0.0 & 8.66 & -4.93 & -7.00 \\
16854.937   & -0.807 &  9.68 & 0.0 &  10.42 & 1.0 & 8.19 & -4.18 & -6.99 \\
16863.004   & -0.776 &  9.99 & 1.0 &  10.72 & 1.0 & 7.92 & -3.34 & -6.75 \\
16888.175*  & -0.636 &  9.69 & 1.0 &  10.42 & 2.0 & 8.20 & -3.52 & -6.99 \\
16888.480*  & -0.973 &  9.69 & 1.0 &  10.42 & 1.0 & 8.20 & -4.18 & -6.99 \\
16890.380   &  0.570 &  9.00 & 2.0 &   9.74 & 3.0 & 0.00 & -5.03 &  0.00 \\
16954.102   & -0.963 &  9.69 & 1.0 &  10.42 & 2.0 & 8.20 & -3.78 & -6.99 \\
16978.063   & -1.198 &  9.69 & 2.0 &  10.42 & 2.0 & 8.23 & -3.52 & -6.99 \\
17323.469   &  0.249 &  9.70 & 3.0 &  10.41 & 4.0 & 0.00 &  0.00 &  0.00 \\
17427.440   & -0.139 &  9.70 & 4.0 &  10.41 & 4.0 & 0.00 &  0.00 &  0.00 \\
17428.140   & -1.694 &  9.70 & 4.0 &  10.41 & 3.0 & 0.00 &  0.00 &  0.00 \\
17448.600   &  0.012 &  9.00 & 2.0 &   9.71 & 1.0 & 0.00 &  0.00 &  0.00 \\
17450.990   & -0.678 &  9.71 & 2.0 &  10.42 & 2.0 & 0.00 &  0.00 &  0.00 \\
17451.320   & -1.033 &  9.71 & 2.0 &  10.42 & 1.0 & 0.00 &  0.00 &  0.00 \\
17455.986   &  0.280 &  9.70 & 2.0 &  10.41 & 3.0 & 0.00 &  0.00 &  0.00 \\
17475.636   & -0.690 &  9.70 & 2.0 &  10.40 & 2.0 & 0.00 &  0.00 &  0.00 \\
17478.020   & -1.636 &  9.71 & 1.0 &  10.42 & 2.0 & 0.00 &  0.00 &  0.00 \\
17505.717   &  0.424 &  9.70 & 3.0 &  10.41 & 4.0 & 0.00 &  0.00 &  0.00 \\
17521.333*  & -0.371 &  9.71 & 2.0 &  10.42 & 2.0 & 0.00 &  0.00 &  0.00 \\
17521.794*  & -1.217 &  9.71 & 2.0 &  10.42 & 3.0 & 0.00 &  0.00 &  0.00 \\
17522.377*  & -2.865 & 10.02 & 2.0 &  10.73 & 2.0 & 0.00 &  0.00 &  0.00 \\
17526.002*  & -1.608 &  9.70 & 3.0 &  10.40 & 2.0 & 0.00 &  0.00 &  0.00 \\
17526.341*  & -0.620 &  9.70 & 3.0 &  10.40 & 3.0 & 0.00 &  0.00 &  0.00 \\
17545.230   & -0.765 &  9.71 & 3.0 &  10.42 & 4.0 & 0.00 &  0.00 &  0.00 \\
17554.006*  & -1.192 &  9.71 & 3.0 &  10.42 & 2.0 & 0.00 &  0.00 &  0.00 \\
17554.471*  & -0.031 &  9.71 & 3.0 &  10.42 & 3.0 & 0.00 &  0.00 &  0.00 \\
17637.468*  &  0.338 &  9.71 & 3.0 &  10.41 & 4.0 & 0.00 &  0.00 &  0.00 \\
17638.184*  & -0.756 &  9.71 & 3.0 &  10.41 & 3.0 & 0.00 &  0.00 &  0.00 \\
17672.044   & -1.974 &  7.95 & 3.0 &   8.65 & 3.0 & 0.00 &  0.00 &  0.00 \\
21023.164   & -0.398 &  9.17 & 0.0 &   9.76 & 1.0 & 0.00 &  0.00 &  0.00 \\
21155.784   & -0.417 &  9.83 & 2.0 &  10.42 & 2.0 & 0.00 &  0.00 &  0.00 \\
21259.256   & -0.400 &  9.83 & 2.0 &  10.42 & 2.0 & 0.00 &  0.00 &  0.00 \\
21259.896   &  0.493 &  9.83 & 2.0 &  10.42 & 3.0 & 0.00 &  0.00 &  0.00 \\
21295.187   & -0.167 &  9.83 & 1.0 &  10.42 & 2.0 & 0.00 &  0.00 &  0.00 \\
22160.413   & -0.430 &  8.77 & 1.0 &   9.33 & 2.0 & 0.00 &  0.00 &  0.00 \\
22166.801   & -0.540 &  8.77 & 1.0 &   9.33 & 1.0 & 0.00 &  0.00 &  0.00 \\
22906.548   & -0.217 &  9.17 & 0.0 &   9.71 & 1.0 & 0.00 &  0.00 &  0.00 \\
\hline                                                                      
Na\,I   &        &       &     &        &     &      &       &       \\
21452.214   & -0.379 &  4.28 & 2.5 &   4.86 & 3.5 & 0.00 &  0.00 &  0.00 \\
22056.399   &  0.287 &  3.19 & 0.5 &   3.75 & 1.5 & 0.00 &  0.00 &  0.00 \\
22059.891   & -1.674 &  4.34 & 1.5 &   4.91 & 0.5 & 0.00 &  0.00 &  0.00 \\
22083.662   & -0.013 &  3.19 & 0.5 &   3.75 & 0.5 & 0.00 &  0.00 &  0.00 \\
23348.378   &  0.281 &  3.75 & 0.5 &   4.28 & 1.5 & 0.00 &  0.00 &  0.00 \\
23378.597   & -0.417 &  3.75 & 1.5 &   4.28 & 1.5 & 0.00 &  0.00 &  0.00 \\
23379.137   &  0.538 &  3.75 & 1.5 &   4.28 & 2.5 & 0.00 &  0.00 &  0.00 \\ 
\hline
\end{tabular}
\end{table*}

\begin{table*}
\addtocounter{table}{-1}
\caption{{\it ...continued.}}
\centering
\begin{tabular}{lrrcrccrr}
\hline
\hline
Wavelength  & $\log(gf)$ & E$_i$i~  & J$_i$  &  E$_f$~    & J$_f$  & $\log(\gamma_{r})$ & $\log(\gamma_{S})$ &  $\log(\gamma_{W})$ \\
~~~~~~~~{\AA} & & eV & & eV & & & & \\
\hline
Mg\,I   &        &       &     &        &     &      &       &       \\
15024.997   &  0.357 &  5.11 & 1.0 &   5.93 & 2.0 & 0.00 &  0.00 &  0.00 \\
15040.246   &  0.135 &  5.11 & 1.0 &   5.93 & 1.0 & 0.00 &  0.00 &  0.00 \\
15748.886*  & -0.338 &  5.93 & 1.0 &   6.72 & 1.0 & 0.00 &  0.00 &  0.00 \\
15748.988*  &  0.140 &  5.93 & 1.0 &   6.72 & 2.0 & 0.00 &  0.00 &  0.00 \\
15765.747*  & -0.337 &  5.93 & 2.0 &   6.72 & 2.0 & 0.00 &  0.00 &  0.00 \\
15765.839*  &  0.411 &  5.93 & 2.0 &   6.72 & 3.0 & 0.00 &  0.00 &  0.00 \\
16632.020*  & -1.840 &  6.73 & 2.0 &   7.47 & 2.0 & 0.00 &  0.00 &  0.00 \\
16632.020*  & -1.080 &  6.73 & 2.0 &   7.47 & 3.0 & 0.00 &  0.00 &  0.00 \\
17108.631   &  0.064 &  5.39 & 0.0 &   6.12 & 1.0 & 0.00 &  0.00 &  0.00 \\
17407.360*  & -0.622 &  6.72 & 3.0 &   7.43 & 4.0 & 0.00 &  0.00 &  0.00 \\
17407.360*  & -0.952 &  6.72 & 1.0 &   7.43 & 2.0 & 0.00 &  0.00 &  0.00 \\
17407.517*  & -0.870 &  6.72 & 2.0 &   7.43 & 3.0 & 0.00 &  0.00 &  0.00 \\
17407.517*  & -2.220 &  6.72 & 2.0 &   7.43 & 3.0 & 0.00 &  0.00 &  0.00 \\
17407.518*  & -1.684 &  6.72 & 3.0 &   7.43 & 2.0 & 0.00 &  0.00 &  0.00 \\
17407.536*  & -0.622 &  6.72 & 1.0 &   7.43 & 4.0 & 0.00 &  0.00 &  0.00 \\
17407.536*  & -0.952 &  6.72 & 2.0 &   7.43 & 2.0 & 0.00 &  0.00 &  0.00 \\
22807.453*  &  0.018 &  6.72 & 3.0 &   7.26 & 4.0 & 0.00 &  0.00 &  0.00 \\
22807.753*  & -1.044 &  6.72 & 3.0 &   7.26 & 3.0 & 0.00 &  0.00 &  0.00 \\
22807.753*  & -2.595 &  6.72 & 3.0 &   7.26 & 2.0 & 0.00 &  0.00 &  0.00 \\
22807.753*  & -1.046 &  6.72 & 2.0 &   7.26 & 2.0 & 0.00 &  0.00 &  0.00 \\
22807.996*  & -1.300 &  6.72 & 3.0 &   7.26 & 3.0 & 0.00 &  0.00 &  0.00 \\
22808.033*  & -1.230 &  6.72 & 2.0 &   7.26 & 3.0 & 0.00 &  0.00 &  0.00 \\
22808.025*  & -0.143 &  6.72 & 2.0 &   7.26 & 3.0 & 0.00 &  0.00 &  0.00 \\
22808.265*  & -0.313 &  6.72 & 1.0 &   7.26 & 2.0 & 0.00 &  0.00 &  0.00 \\
\hline                                                                      
Mg\,II  &        &       &     &        &     &      &       &       \\
16760.219   &  0.471 & 12.08 & 0.5 &1 12.82 & 1.5 & 7.83 & -4.26 & -7.00 \\
16799.931   &  0.730 & 12.08 & 1.5 &1 12.82 & 2.5 & 0.00 &  0.00 &  0.00 \\
\hline                                                                      
Al\,I   &        &       &     &        &     &      &       &       \\
16718.957   &  0.152 &  4.09 & 0.5 &   4.83 & 1.5 & 7.74 & -4.69 & -7.52 \\
16750.564   &  0.408 &  4.09 & 1.5 &   4.83 & 2.5 & 7.74 & -4.69 & -7.52 \\
16763.396   & -0.480 &  4.09 & 1.5 &   4.83 & 1.5 & 7.56 & -4.73 & -7.08 \\
21093.029   & -0.309 &  4.09 & 0.5 &   4.67 & 0.5 & 7.84 & -4.74 & -7.45 \\
21163.755   & -0.009 &  4.09 & 1.5 &   4.67 & 0.5 & 7.84 & -4.74 & -7.45 \\
\hline                                                                      
Si\,I       &        &       &     &        &     &      &       &       \\
15376.831   & -0.691 &  6.22 & 2.0 &   7.03 & 2.0 & 0.00 &  0.00 &  0.00 \\
15960.063   &  0.087 &  5.98 & 3.0 &   6.76 & 2.0 & 0.00 & -4.74 &  0.00 \\
16060.009   & -0.484 &  5.95 & 1.0 &   6.73 & 0.0 & 0.00 &  0.00 &  0.00 \\
16094.787   & -0.144 &  5.96 & 2.0 &   6.73 & 1.0 & 0.00 &  0.00 &  0.00 \\
16163.691   & -0.862 &  5.95 & 1.0 &   6.72 & 2.0 & 8.58 & -5.15 & -7.10 \\
16215.670   & -0.631 &  5.95 & 1.0 &   6.72 & 1.0 & 7.45 & -3.01 & -6.98 \\
16241.833   & -0.770 &  5.96 & 2.0 &   6.73 & 3.0 & 8.58 & -5.16 & -7.44 \\
16346.857   & -1.088 &  7.11 & 2.0 &   7.87 & 3.0 & 7.17 & -2.67 & -6.86 \\
16412.982   & -0.729 &  6.80 & 1.0 &   7.56 & 2.0 & 8.19 & -3.95 & -6.75 \\
16434.927   & -1.187 &  5.96 & 2.0 &   6.72 & 1.0 & 8.57 & -5.16 & -7.10 \\
16680.770   & -0.141 &  5.98 & 3.0 &   6.73 & 3.0 & 0.00 &  0.00 &  0.00 \\
16828.160   &        &       &     &         &     &      &       &      \\
17225.570   &  0.116 &  6.62 & 1.0 &   7.34 & 2.0 & 0.00 &  0.00 &  0.00 \\
17327.275   &  0.698 &  6.62 & 3.0 &   7.33 & 4.0 & 0.00 &  0.00 &  0.00 \\
17466.879   &  0.101 &  6.62 & 1.0 &   7.33 & 2.0 & 0.00 &  0.00 &  0.00 \\
17616.946   &  0.130 &  6.62 & 3.0 &   7.32 & 4.0 & 0.00 &  0.00 &  0.00 \\
17623.328   & -0.624 &  6.62 & 3.0 &   7.32 & 3.0 & 0.00 &  0.00 &  0.00 \\
20602.832   &  0.106 &  6.73 & 3.0 &   7.33 & 3.0 & 0.00 &  0.00 &  0.00 \\
20804.157   & -1.021 &  6.12 & 1.0 &   6.72 & 2.0 & 8.58 & -5.08 & -7.44 \\
20827.496   & -1.180 &  7.01 & 2.0 &   7.60 & 3.0 & 7.54 & -3.49 & -7.13 \\
20917.152   &  0.575 &  6.73 & 3.0 &   7.32 & 4.0 & 0.00 &  0.00 &  0.00 \\
20926.149   & -1.074 &  6.73 & 3.0 &   7.32 & 3.0 & 8.60 & -3.63 & -7.33 \\
21354.199   &  0.193 &  6.22 & 2.0 &   6.80 & 1.0 & 0.00 &  0.00 &  0.00 \\
21779.660   &  0.418 &  6.72 & 1.0 &   7.29 & 2.0 & 0.00 &  0.00 &  0.00 \\
21819.671   &  0.168 &  6.72 & 2.0 &   7.29 & 3.0 & 0.00 &  0.00 &  0.00 \\
21874.140   & -0.599 &  6.72 & 2.0 &   7.29 & 2.0 & 0.00 &  0.00 &  0.00 \\
\hline
\end{tabular}
\end{table*}

\begin{table*}
\addtocounter{table}{-1}
\caption{{\it ...continued.}}
\centering
\begin{tabular}{lrrcrccrr}
\hline
\hline
Wavelength  & $\log(gf)$ & E$_i$i~  & J$_i$  &  E$_f$~    & J$_f$  & $\log(\gamma_{r})$ & $\log(\gamma_{S})$ &  $\log(\gamma_{W})$ \\
~~~~~~~~{\AA} & & eV & & eV & & & & \\
\hline
21879.324   &  0.406 &  6.72 & 2.0 &   7.29 & 3.0 & 0.00 &  0.00 &  0.00 \\
22062.710   &  0.538 &  6.73 & 3.0 &   7.29 & 4.0 & 0.00 &  0.00 &  0.00 \\
22537.534   & -0.226 &  6.62 & 3.0 &   7.17 & 2.0 & 0.00 &  0.00 &  0.00 \\
22665.757   & -0.680 &  6.62 & 1.0 &   7.17 & 2.0 & 0.00 &  0.00 &  0.00 \\
\hline                                                         
P\,I    &        &       &     &        &     &      &       &       \\
16482.932   & -0.310 &  7.21 & 1.5 &   7.96 & 0.5 & 8.77 & -5.59 & -7.27 \\
\hline                                                                      
S\,I    &        &       &     &        &     &      &       &       \\
15402.331   &  0.350 &  8.70 & 1.0 &   9.50 & 2.0 & 0.00 &  0.00 &  0.00 \\
15403.724*  & -0.180 &  8.70 & 1.0 &   9.50 & 2.0 & 8.61 & -4.82 & -7.32 \\
15403.791*  &  0.730 &  8.70 & 2.0 &   9.50 & 3.0 & 0.00 &  0.00 &  0.00 \\
15405.979   & -0.380 &  8.70 & 2.0 &   9.50 & 2.0 & 0.00 &  0.00 &  0.00 \\
15422.276   &  0.890 &  8.70 & 3.0 &   9.50 & 4.0 & 8.62 & -4.81 & -7.32 \\
15469.816   & -0.050 &  8.05 & 1.0 &   8.85 & 1.0 & 8.44 & -4.94 & -7.41 \\
15475.616   & -0.520 &  8.05 & 0.0 &   8.85 & 1.0 & 8.44 & -4.94 & -7.41 \\
15478.482   &  0.180 &  8.05 & 2.0 &   8.85 & 1.0 & 8.44 & -4.94 & -7.41 \\
16464.375   & -0.951 &  8.41 & 3.0 &   9.16 & 2.0 & 8.35 & -4.60 & -7.00 \\
16466.740   & -0.382 &  9.17 & 3.0 &   9.92 & 4.0 & 7.14 & -3.94 & -6.67 \\
16542.660   & -0.070 &  8.42 & 4.0 &   9.17 & 3.0 & 7.10 & -4.60 & -7.35 \\
16576.621   & -0.850 &  8.42 & 3.0 &   9.17 & 3.0 & 7.89 & -4.60 & -7.30 \\
16580.887   & -1.113 &  8.42 & 1.0 &   9.16 & 2.0 & 7.32 & -4.60 & -7.00 \\
16587.133*  & -0.550 &  8.42 & 2.0 &   9.16 & 2.0 & 7.58 & -4.60 & -7.00 \\
16587.194*  & -0.975 &  8.42 & 0.0 &   9.16 & 1.0 & 7.10 & -4.60 & -7.00 \\
16587.249*  & -1.010 &  8.42 & 2.0 &   9.16 & 2.0 & 0.00 &  0.00 &  0.00 \\
16593.220   & -0.560 &  8.42 & 3.0 &   9.16 & 2.0 & 7.89 & -4.60 & -7.35 \\
22507.344   & -0.480 &  7.87 & 1.0 &   8.42 & 2.0 & 0.00 &  0.00 &  0.00 \\
22519.066   & -0.250 &  7.87 & 1.0 &   8.42 & 1.0 & 0.00 &  0.00 &  0.00 \\
22526.053   & -0.510 &  7.87 & 1.0 &   8.42 & 0.0 & 0.00 &  0.00 &  0.00 \\
22563.828   & -0.260 &  7.87 & 2.0 &   8.42 & 2.0 & 0.00 &  0.00 &  0.00 \\
22575.394   & -0.730 &  7.87 & 2.0 &   8.42 & 1.0 & 0.00 &  0.00 &  0.00 \\
22644.056   & -0.340 &  7.87 & 3.0 &   8.42 & 3.0 & 0.00 &  0.00 &  0.00 \\
22707.736   &  0.440 &  7.87 & 3.0 &   8.42 & 4.0 & 0.00 &  0.00 &  0.00 \\
\hline                                                                      
K\,I    &        &       &     &        &     &      &       &       \\
15163.069*  & -0.613 &  2.67 & 2.5 &   3.49 & 2.5 & 7.62 & -4.36 &  0.00 \\
15163.069*  &  0.689 &  2.67 & 2.5 &   3.49 & 3.5 & 7.62 & -4.36 &  0.00 \\
15168.376   &  0.480 &  2.67 & 1.5 &   3.49 & 2.5 & 7.62 & -4.36 &  0.00 \\
\hline                                                                      
Ca\,I   &        &       &     &        &     &      &       &       \\
16136.823   & -0.363 &  5.30 & 1.0 &   4.53 & 0.0 & 7.55 & -4.38 & -7.36 \\
16150.763   & -0.032 &  5.30 & 2.0 &   4.53 & 1.0 & 7.56 & -4.38 & -7.36 \\
16197.075   &  0.254 &  5.30 & 3.0 &   4.53 & 2.0 & 7.55 & -4.37 & -7.36 \\
16204.087   &        &       &     &        &     &      &       &      \\
20937.904   & -1.453 &  4.68 & 2.0 &   5.27 & 2.0 & 8.06 & -4.15 & -7.30 \\
20962.410   & -0.703 &  4.68 & 3.0 &   5.27 & 2.0 & 8.06 & -4.15 & -7.30 \\
22607.945   &  0.516 &  4.68 & 1.0 &   5.23 & 2.0 & 8.20 & -5.27 & -7.59 \\
22624.962   &  0.687 &  4.68 & 2.0 &   5.23 & 3.0 & 8.20 & -5.27 & -7.59 \\
22651.178   &  0.847 &  4.68 & 3.0 &   5.23 & 4.0 & 8.20 & -5.27 & -7.59 \\
22821.065   & -0.396 &  4.62 & 2.0 &   5.17 & 1.0 & 8.08 & -4.28 & -7.35 \\
\hline                                                                      
Ca\,II  &        &       &     &        &     &      &       &       \\
16649.877   &  0.642 &  9.98 & 2.5 &   9.24 & 1.5 & 7.98 & -4.58 & -7.28 \\
21389.021*  &  0.177 &  9.02 & 2.5 &   8.44 & 3.5 & 8.70 & -5.05 & -7.54 \\
21389.021*  & -1.124 &  9.02 & 2.5 &   8.44 & 2.5 & 8.70 & -5.05 & -7.54 \\
21428.908   &  0.021 &  9.02 & 1.5 &   8.44 & 2.5 & 8.70 & -5.05 & -7.54 \\
\hline                                                                      
Ti\,II  &        &       &     &        &     &      &       &       \\
15873.838   & -1.814 &  3.12 & 2.5 &   3.90 & 3.5 & 8.16 & -6.33 & -7.50 \\
16005.054   & -2.070 &  3.09 & 1.5 &   3.87 & 2.5 & 8.15 & -6.33 & -7.50 \\
16621.818   & -2.546 &  3.12 & 2.5 &   3.87 & 2.5 & 8.15 & -6.33 & -7.85 \\
\hline                                                                      
Mn\,I   &        &       &     &        &     &      &       &       \\
15217.686   &  0.520 &  4.89 & 3.5 &   5.70 & 3.5 & 8.03 & -4.93 & -7.52 \\
\hline                
\end{tabular}
\end{table*}

\begin{table*}
\addtocounter{table}{-1}
\caption{{\it ...continued.}}
\centering
\begin{tabular}{lrrcrccrr}
\hline
\hline
Wavelength  & $\log(gf)$ & E$_i$i~  & J$_i$  &  E$_f$~    & J$_f$  & $\log(\gamma_{r})$ & $\log(\gamma_{S})$ &  $\log(\gamma_{W})$ \\
~~~~~~~~{\AA} & & eV & & eV & & & & \\
\hline                                                                                         Fe\,I   &        &       &     &        &     &      &       &       \\
15051.749   &  0.426 &  5.35 & 4.0 &   6.18 & 3.0 & 8.22 & -4.86 & -7.49 \\
15077.287   & -0.244 &  5.59 & 3.0 &   6.41 & 3.0 & 8.32 & -5.00 & -7.49 \\
15094.695   &  0.603 &  6.36 & 3.0 &   7.18 & 4.0 & 8.24 & -4.75 & -7.46 \\
15144.051   & -0.401 &  5.64 & 1.0 &   6.46 & 1.0 & 8.29 & -3.91 & -7.29 \\
15207.526   &  0.323 &  5.39 & 3.0 &   6.20 & 2.0 & 8.22 & -4.86 & -7.49 \\
15219.618   & -0.825 &  5.59 & 3.0 &   6.40 & 2.0 & 8.18 & -4.42 & -7.45 \\
15224.729   & -0.315 &  5.96 & 2.0 &   6.77 & 1.0 & 8.24 & -4.63 & -7.44 \\
15239.712   & -0.032 &  6.42 & 3.0 &   7.23 & 2.0 & 8.16 & -4.60 & -7.35 \\
15244.973   & -0.072 &  5.59 & 3.0 &   6.40 & 3.0 & 8.18 & -4.50 & -7.45 \\
15381.960   & -0.461 &  5.98 & 1.0 &   6.78 & 0.0 & 8.26 & -4.65 & -7.08 \\
15387.803   & -0.184 &  7.08 & 4.0 &   6.28 & 3.0 & 7.78 & -4.30 & -7.29 \\
15490.529*  & -0.340 &  6.29 & 2.0 &   7.09 & 2.0 & 8.23 & -4.15 & -6.98 \\
15490.884*  & -0.466 &  6.29 & 2.0 &   7.09 & 1.0 & 8.23 & -4.43 & -7.32 \\
15514.279   & -0.473 &  6.29 & 5.0 &   7.09 & 5.0 & 8.13 & -4.32 & -7.32 \\
15537.697   & -0.032 &  6.32 & 2.0 &   7.12 & 3.0 & 8.35 & -4.41 & -6.98 \\
15542.079   & -0.337 &  5.64 & 1.0 &   6.44 & 0.0 & 8.16 & -5.18 & -7.45 \\
15588.259   &  0.419 &  6.37 & 4.0 &   7.16 & 4.0 & 8.07 & -4.65 & -7.34 \\
15591.490   &  0.874 &  6.24 & 6.0 &   7.04 & 7.0 & 8.39 & -4.42 & -7.33 \\
15591.497   & -0.687 &  6.36 & 2.0 &   7.16 & 3.0 & 8.36 & -4.47 & -7.32 \\
15594.399   & -0.091 &  6.35 & 1.0 &   7.14 & 2.0 & 8.41 & -4.62 & -6.99 \\
15598.772   & -0.281 &  6.35 & 1.0 &   7.14 & 1.0 & 8.41 & -4.53 & -6.99 \\
15604.220   &  0.538 &  6.24 & 6.0 &   7.04 & 6.0 & 8.39 & -4.47 & -7.33 \\
15611.146   & -3.768 &  3.41 & 1.0 &   4.21 & 2.0 & 6.88 & -6.13 & -7.79 \\
15621.654   &  0.589 &  5.54 & 4.0 &   6.33 & 4.0 & 8.14 & -4.36 & -7.45 \\
15631.948   &  0.120 &  5.35 & 4.0 &   6.14 & 4.0 & 8.22 & -4.85 & -7.49 \\
15648.510   & -0.599 &  5.43 & 1.0 &   6.22 & 1.0 & 8.22 & -4.87 & -7.49 \\
15652.871   & -0.161 &  6.25 & 5.0 &   7.04 & 4.0 & 8.41 & -4.48 & -7.32 \\
15662.013   &  0.368 &  5.83 & 5.0 &   6.62 & 4.0 & 8.18 & -5.32 & -7.11 \\
15665.240   & -0.337 &  5.98 & 1.0 &   6.77 & 1.0 & 8.24 & -4.63 & -7.44 \\
15671.003   & -0.216 &  6.33 & 1.0 &   7.12 & 2.0 & 8.33 & -4.68 & -6.98 \\
15723.586   & -0.143 &  5.62 & 2.0 &   6.41 & 3.0 & 8.32 & -5.00 & -7.01 \\
15761.313   &  0.143 &  6.25 & 4.0 &   7.04 & 3.0 & 8.28 & -4.45 & -6.98 \\
15774.068   &  0.545 &  6.30 & 4.0 &   7.09 & 5.0 & 8.30 & -3.71 & -7.26 \\
15788.997   &  0.494 &  6.25 & 4.0 &   7.04 & 4.0 & 8.28 & -4.54 & -6.98 \\
15892.395   &  0.004 &  6.31 & 2.0 &   7.09 & 2.0 & 8.30 & -4.15 & -6.98 \\
15892.452   & -0.142 &  6.34 & 2.0 &   7.12 & 3.0 & 8.20 & -4.82 & -6.98 \\
15892.769   &  0.124 &  6.31 & 2.0 &   7.09 & 1.0 & 8.30 & -4.43 & -6.98 \\
15934.017   & -0.298 &  6.31 & 1.0 &   7.09 & 2.0 & 8.20 & -4.17 & -6.98 \\
15938.918   &  0.068 &  6.37 & 2.0 &   7.15 & 3.0 & 8.17 & -4.47 & -6.98 \\
15980.726   &  0.958 &  6.26 & 6.0 &   7.04 & 7.0 & 8.13 & -4.68 & -6.98 \\
16006.758   &  0.747 &  6.35 & 4.0 &   7.12 & 5.0 & 8.37 & -4.49 & -6.98 \\
16007.083   &  0.218 &  6.35 & 4.0 &   7.12 & 4.0 & 8.38 & -4.51 & -6.99 \\
16008.075   &  0.279 &  6.27 & 3.0 &   7.04 & 2.0 & 8.31 & -4.37 & -6.98 \\
16009.611   & -0.480 &  5.43 & 1.0 &   6.20 & 2.0 & 8.22 & -4.86 & -7.15 \\
16037.819   &  0.253 &  6.26 & 6.0 &   7.04 & 6.0 & 8.15 & -4.47 & -6.98 \\
16039.854   & -0.842 &  5.65 & 0.0 &   6.43 & 1.0 & 8.17 & -4.47 & -7.11 \\
16040.654   &  0.313 &  5.87 & 4.0 &   6.65 & 4.0 & 8.18 & -4.69 & -7.11 \\
16042.716   &  0.353 &  6.27 & 3.0 &   7.04 & 3.0 & 8.31 & -4.45 & -6.98 \\
16174.976   &  0.188 &  6.38 & 3.0 &   7.15 & 4.0 & 8.38 & -4.46 & -6.98 \\
16179.583   &  0.140 &  6.32 & 6.0 &   7.09 & 6.0 & 8.38 & -4.40 & -6.98 \\
16180.900   &  0.296 &  6.27 & 3.0 &   7.04 & 2.0 & 8.23 & -4.43 & -6.98 \\
16195.060   &  0.485 &  6.39 & 3.0 &   7.16 & 4.0 & 8.35 & -4.52 & -6.99 \\
16204.253   &  0.206 &  6.32 & 4.0 &   7.09 & 4.0 & 8.15 & -3.98 & -6.98 \\
16207.744   &  0.585 &  6.32 & 4.0 &   7.09 & 5.0 & 8.14 & -4.32 & -7.32 \\
16213.537   &  0.321 &  6.27 & 3.0 &   7.04 & 3.0 & 8.23 & -4.49 & -7.32 \\
16225.618   &  0.295 &  6.38 & 3.0 &   7.14 & 4.0 & 8.38 & -4.58 & -7.32 \\
\hline                                    
\end{tabular}
\end{table*}
                                
\begin{table*}
\addtocounter{table}{-1}
\caption{{\it ...continued.}}
\centering
\begin{tabular}{lrrcrccrr}
\hline
\hline
Wavelength  & $\log(gf)$ & E$_i$i~  & J$_i$  &  E$_f$~    & J$_f$  & $\log(\gamma_{r})$ & $\log(\gamma_{S})$ &  $\log(\gamma_{W})$ \\
~~~~~~~~{\AA} & & eV & & eV & & & & \\
\hline                                                         

16231.645   &  0.600 &  6.38 & 4.0 &   7.14 & 5.0 & 8.09 & -4.50 & -7.33 \\
16246.460   &  0.062 &  6.27 & 3.0 &   7.04 & 4.0 & 8.23 & -4.51 & -6.98 \\
16252.550   & -0.270 &  6.32 & 2.0 &   7.09 & 3.0 & 8.35 & -4.03 & -6.98 \\
16284.769   & -0.175 &  6.40 & 3.0 &   7.16 & 4.0 & 8.06 & -4.30 & -7.32 \\
16292.840   & -0.157 &  5.92 & 3.0 &   6.68 & 3.0 & 8.12 & -5.37 & -7.54 \\
16316.320   &  1.073 &  6.28 & 7.0 &   7.04 & 8.0 & 8.38 & -4.35 & -7.32 \\
16324.451   & -0.556 &  5.39 & 3.0 &   6.14 & 4.0 & 8.22 & -4.85 & -7.52 \\
16331.523   & -0.424 &  5.98 & 1.0 &   6.74 & 2.0 & 8.17 & -4.50 & -7.12 \\
16333.141   & -0.774 &  5.64 & 1.0 &   6.40 & 2.0 & 8.18 & -4.50 & -7.11 \\
16394.388   &  0.358 &  5.96 & 2.0 &   6.71 & 3.0 & 8.04 & -4.34 & -7.47 \\
16398.167   &  0.111 &  5.92 & 3.0 &   6.68 & 4.0 & 8.13 & -4.29 & -7.54 \\
16436.621   &  0.028 &  5.92 & 3.0 &   6.68 & 3.0 & 8.13 & -4.51 & -7.12 \\
16440.394   & -0.241 &  6.29 & 2.0 &   7.04 & 2.0 & 8.42 & -4.37 & -7.33 \\
16444.816   &  0.663 &  5.83 & 5.0 &   6.58 & 5.0 & 8.10 & -4.30 & -7.45 \\
16466.923   & -0.107 &  6.39 & 3.0 &   7.15 & 4.0 & 8.35 & -4.46 & -6.98 \\
16471.754   &  0.026 &  6.37 & 2.0 &   7.12 & 3.0 & 8.18 & -4.82 & -6.98 \\
16481.228   & -0.165 &  6.39 & 1.0 &   7.15 & 2.0 & 8.39 & -4.53 & -6.98 \\
16486.667   &  0.783 &  5.83 & 5.0 &   6.58 & 6.0 & 8.00 & -4.86 & -7.11 \\
16506.293   & -0.457 &  5.95 & 4.0 &   6.70 & 4.0 & 8.06 & -4.73 & -7.14 \\
16517.223   &  0.674 &  6.29 & 5.0 &   7.04 & 6.0 & 8.14 & -4.47 & -6.98 \\
16522.074   &  0.319 &  6.29 & 5.0 &   7.04 & 5.0 & 8.14 & -4.93 & -6.98 \\
16524.467   &  0.679 &  6.34 & 5.0 &   7.09 & 6.0 & 8.24 & -4.35 & -6.98 \\
16531.983   & -0.823 &  6.29 & 5.0 &   7.04 & 5.0 & 8.13 & -4.46 & -7.33 \\
16539.193   & -0.128 &  6.34 & 5.0 &   7.09 & 5.0 & 8.24 & -4.32 & -6.98 \\
16544.667   & -0.030 &  6.34 & 5.0 &   7.09 & 5.0 & 8.24 & -4.39 & -6.99 \\
16551.995   &  0.336 &  6.41 & 2.0 &   7.16 & 3.0 & 8.02 & -4.74 & -6.98 \\
16559.677   &  0.263 &  6.40 & 3.0 &   7.15 & 4.0 & 8.08 & -4.36 & -6.98 \\
16561.764   &  0.243 &  5.98 & 1.0 &   6.73 & 2.0 & 8.03 & -4.53 & -7.46 \\
16581.383   & -0.680 &  5.92 & 3.0 &   6.67 & 3.0 & 8.20 & -4.76 & -7.55 \\
16612.764   &  0.286 &  6.40 & 3.0 &   7.14 & 4.0 & 8.06 & -4.30 & -7.32 \\
16632.504   & -0.228 &  5.87 & 4.0 &   6.62 & 4.0 & 8.18 & -5.28 & -7.11 \\
16645.877   & -0.032 &  5.96 & 2.0 &   6.70 & 2.0 & 8.07 & -4.58 & -7.26 \\
16665.484   & -0.039 &  6.02 & 3.0 &   6.76 & 2.0 & 8.19 & -4.72 & -7.47 \\
16723.311   & -0.523 &  5.92 & 3.0 &   6.66 & 3.0 & 8.12 & -4.76 & -6.52 \\
16753.067   &  0.407 &  6.38 & 4.0 &   7.12 & 5.0 & 8.09 & -4.57 & -7.32 \\
16760.148   &  0.311 &  6.61 & 4.0 &   7.35 & 5.0 & 7.92 & -3.58 & -6.92 \\
16856.442   & -0.062 &  6.35 & 3.0 &   7.09 & 3.0 & 8.19 & -4.03 & -6.98 \\
16857.136   & -0.214 &  5.95 & 4.0 &   6.68 & 3.0 & 8.09 & -5.38 & -7.20 \\
16858.523   &  0.133 &  6.35 & 3.0 &   7.09 & 4.0 & 8.19 & -3.98 & -6.98 \\
17005.449   &  0.005 &  6.07 & 2.0 &   6.80 & 1.0 & 8.11 & -4.93 & -7.47 \\
17011.095   &  0.102 &  5.95 & 4.0 &   6.68 & 3.0 & 8.10 & -4.57 & -7.48 \\
17161.103   & -0.071 &  6.02 & 3.0 &   6.74 & 3.0 & 8.15 & -5.29 & -7.34 \\
21238.466   & -1.423 &  5.54 & 4.0 &   4.96 & 3.0 & 8.21 & -4.71 & -7.29 \\
22257.107   & -0.710 &  5.62 & 2.0 &   5.06 & 3.0 & 8.84 & -5.41 & -7.54 \\
22260.180   & -0.941 &  5.64 & 1.0 &   5.09 & 2.0 & 8.84 & -5.38 & -7.54 \\
22380.797   & -0.464 &  5.59 & 3.0 &   5.03 & 4.0 & 8.84 & -5.38 & -7.75 \\
22419.976   & -0.155 &  6.77 & 2.0 &   6.22 & 1.0 & 7.49 & -4.79 & -7.34 \\
22473.263   &  0.442 &  6.67 & 5.0 &   6.12 & 5.0 & 7.50 & -4.78 & -7.34 \\
22619.837   & -0.362 &  5.54 & 4.0 &   4.99 & 5.0 & 8.85 & -5.41 & -7.54 \\
22740.416   & -0.016 &  6.75 & 3.0 &   6.20 & 2.0 & 7.67 & -4.86 & -7.36 \\
22832.364   & -1.269 &  5.64 & 1.0 &   5.10 & 1.0 & 8.84 & -5.38 & -7.54 \\
22845.940   & -0.673 &  5.83 & 5.0 &   6.37 & 4.0 & 8.03 & -4.67 & -7.45 \\
23144.581   & -0.069 &  6.71 & 4.0 &   6.18 & 3.0 & 7.51 & -4.80 & -7.34 \\
23566.659   &  0.306 &  6.67 & 5.0 &   6.14 & 4.0 & 7.50 & -4.78 & -7.34 \\
\hline                         
\end{tabular}
\end{table*}
                                
\begin{table*}
\addtocounter{table}{-1}
\caption{{\it ...continued.}}
\centering
\begin{tabular}{lrrcrccrr}
\hline
\hline
Wavelength  & $\log(gf)$ & E$_i$i~  & J$_i$  &  E$_f$~    & J$_f$  & $\log(\gamma_{r})$ & $\log(\gamma_{S})$ &  $\log(\gamma_{W})$ \\
~~~~~~~~{\AA} & & eV & & eV & & & & \\
\hline                                                         
Ni\,I   &        &       &     &        &     &      &       &       \\
16310.480*  &  0.055 &  5.28 & 3.0 &   6.04 & 2.0 & 8.03 & -4.59 & -7.49 \\
16310.533*  & -0.359 &  5.28 & 3.0 &   6.04 & 2.0 & 8.03 & -4.59 & -7.49 \\
16310.562*  & -1.721 &  5.28 & 3.0 &   6.04 & 2.0 & 8.03 & -4.59 & -7.49 \\
16310.590*  & -1.217 &  5.28 & 3.0 &   6.04 & 2.0 & 8.03 & -4.59 & -7.49 \\
16310.638*  & -1.811 &  5.28 & 3.0 &   6.04 & 2.0 & 8.03 & -4.59 & -7.49 \\
16363.104   &  0.588 &  5.28 & 3.0 &   6.04 & 4.0 & 7.93 & -4.64 & -7.14 \\
16364.188   &  0.442 &  5.28 & 3.0 &   6.04 & 4.0 & 8.41 & -4.90 & -7.54 \\
16867.283   &  0.115 &  5.47 & 1.0 &   6.20 & 2.0 & 7.92 & -4.81 & -7.17 \\
16996.261*  &  0.291 &  5.31 & 2.0 &   6.03 & 3.0 & 7.93 & -4.66 & -7.47 \\
16996.268*  & -0.123 &  5.31 & 2.0 &   6.03 & 3.0 & 7.93 & -4.66 & -7.47 \\
16996.270*  & -1.485 &  5.31 & 2.0 &   6.03 & 3.0 & 7.93 & -4.66 & -7.47 \\
16996.271*  & -0.981 &  5.31 & 2.0 &   6.03 & 3.0 & 7.93 & -4.66 & -7.47 \\
16996.273*  & -1.575 &  5.31 & 2.0 &   6.03 & 3.0 & 7.93 & -4.66 & -7.47 \\
17001.025   &  0.379 &  5.49 & 2.0 &   6.22 & 3.0 & 7.99 & -4.76 & -7.50 \\
\hline                                                                      
Ce\,III &        &       &     &        &     &      &       &       \\
15847.581   & -0.838 &  0.19 & 5.0 &   0.97 & 4.0 & 0.00 &  0.00 & -7.08 \\
15956.790   & -0.926 &  0.00 & 4.0 &   0.78 & 3.0 & 0.00 &  0.00 & -7.08 \\
\hline                                                                      
Yb\,II  &        &       &     &        &     &      &       &       \\
16498.420   & -0.640 &  3.02 & 2.5 &   3.77 & 1.5 & 0.00 &  0.00 & -7.06 \\ 
\hline
\end{tabular}    
   \end{table*}

\begin{sidewaystable*}[ht]
\caption[]{LTE abundances for 16 chemical elements derived from the IGRINS spectra for the 23 classical Cepheids. Abundances are given as [X/H] relative to the solar values \citep{grevesse2011chemical}, along with associated uncertainties. The table is divided into two parts for clarity, with the first part covering elements from C to S and the second part from K to Yb.}
\label{abund}
\centering
\begin{tabular}{lrrrrrrrr}
\hline
\hline
ID                &     [C/H]~~~~       &    [N/H]~~~~     &    [Na/H]~~~~     &     [Mg/H]~~~~     &     [Al/H]~~~~     &   [Si/H]~~~~       &     [P/H]~~~~      &    [S/H]~~~~           \\
\hline
AQ Car            &  $-$0.31~$\pm$~0.20 & 0.41~$\pm$~0.15  &   0.32~$\pm$~0.20 &    0.21~$\pm$~0.19 &    0.26~$\pm$~0.05 &    0.13~$\pm$~0.19 & $-$0.06~$\pm$~0.15 &    0.00~$\pm$~0.16    \\
AQ Pup            &  $-$0.34~$\pm$~0.11 & 0.22~$\pm$~0.15  &   0.35~$\pm$~0.20 &    0.12~$\pm$~0.14 &    0.11~$\pm$~0.04 & $-$0.09~$\pm$~0.14 & $-$0.06~$\pm$~0.15 & $-$0.05~$\pm$~0.20    \\
DR Vel            &  $-$0.20~$\pm$~0.07 & 0.28~$\pm$~0.15  &   0.35~$\pm$~0.18 &    0.00~$\pm$~0.14 & $-$0.05~$\pm$~0.20 & $-$0.14~$\pm$~0.11 & $-$0.03~$\pm$~0.15 & $-$0.04~$\pm$~0.18    \\
RV Sco            &  $-$0.42~$\pm$~0.15 & 0.48~$\pm$~0.15  &   0.29~$\pm$~0.17 &    0.10~$\pm$~0.11 &    0.13~$\pm$~0.01 & $-$0.05~$\pm$~0.17 & $-$0.05~$\pm$~0.15 & $-$0.00~$\pm$~0.16    \\
S Nor             &  $-$0.26~$\pm$~0.13 & 0.37~$\pm$~0.15  &   0.34~$\pm$~0.20 &    0.07~$\pm$~0.09 &    0.10~$\pm$~0.10 &    0.00~$\pm$~0.14 & $-$0.03~$\pm$~0.15 & $-$0.01~$\pm$~0.20    \\
SS Cma            &  $-$0.35~$\pm$~0.18 & 0.42~$\pm$~0.15  &   0.29~$\pm$~0.20 &    0.05~$\pm$~0.17 &    0.12~$\pm$~0.10 &    0.06~$\pm$~0.15 & $-$0.05~$\pm$~0.15 &    0.03~$\pm$~0.15    \\
S Tra             &  $-$0.20~$\pm$~0.20 & 0.28~$\pm$~0.15  &   0.20~$\pm$~0.12 &    0.25~$\pm$~0.20 &    0.19~$\pm$~0.07 & $-$0.02~$\pm$~0.16 & $-$0.02~$\pm$~0.15 &    0.05~$\pm$~0.20    \\
SV Vel            &  $-$0.39~$\pm$~0.09 & 0.48~$\pm$~0.15  &   0.33~$\pm$~0.20 &    0.02~$\pm$~0.12 & $-$0.03~$\pm$~0.07 & $-$0.15~$\pm$~0.14 & $-$0.03~$\pm$~0.15 & $-$0.12~$\pm$~0.16    \\
SV Vul            &     0.12~$\pm$~0.15 & 0.38~$\pm$~0.15  &   0.20~$\pm$~0.15 &    0.23~$\pm$~0.18 &    0.07~$\pm$~0.08 &    0.03~$\pm$~0.14 &    0.07~$\pm$~0.15 &    0.12~$\pm$~0.15    \\
T Mon             &  $-$0.25~$\pm$~0.19 & 0.56~$\pm$~0.15  &   0.55~$\pm$~0.15 &    0.20~$\pm$~0.12 &    0.09~$\pm$~0.07 & $-$0.06~$\pm$~0.12 &    0.19~$\pm$~0.15 &    0.25~$\pm$~0.20    \\
U Aql             &  $-$0.20~$\pm$~0.18 & 0.43~$\pm$~0.15  &   0.30~$\pm$~0.16 &    0.03~$\pm$~0.20 &    0.16~$\pm$~0.10 &    0.10~$\pm$~0.20 & $-$0.01~$\pm$~0.15 &    0.06~$\pm$~0.20    \\
U Car             &  $-$0.14~$\pm$~0.17 & 0.51~$\pm$~0.15  &   0.50~$\pm$~0.20 &    0.14~$\pm$~0.14 &    0.48~$\pm$~0.07 &    0.20~$\pm$~0.14 &    0.10~$\pm$~0.15 &    0.05~$\pm$~0.19    \\
U Sgr             &  $-$0.21~$\pm$~0.17 & 0.39~$\pm$~0.15  &   0.41~$\pm$~0.20 &    0.23~$\pm$~0.17 &    0.22~$\pm$~0.11 &    0.23~$\pm$~0.17 &    0.19~$\pm$~0.15 &    0.15~$\pm$~0.20    \\
V0636 Sco         &  $-$0.17~$\pm$~0.12 & 0.35~$\pm$~0.15  &   0.37~$\pm$~0.20 &    0.16~$\pm$~0.10 &    0.09~$\pm$~0.17 &    0.02~$\pm$~0.14 & $-$0.04~$\pm$~0.15 & $-$0.10~$\pm$~0.17    \\
VZ Pup            &  $-$0.57~$\pm$~0.12 & 0.11~$\pm$~0.15  &   0.30~$\pm$~0.17 &    0.14~$\pm$~0.18 &    0.11~$\pm$~0.14 & $-$0.15~$\pm$~0.18 & $-$0.08~$\pm$~0.15 & $-$0.05~$\pm$~0.20    \\
W Gem             &  $-$0.37~$\pm$~0.14 & 0.30~$\pm$~0.15  &   0.29~$\pm$~0.17 &    0.10~$\pm$~0.20 &    0.10~$\pm$~0.12 &    0.14~$\pm$~0.12 & $-$0.04~$\pm$~0.15 &    0.03~$\pm$~0.20    \\
WX Pup            &  $-$0.31~$\pm$~0.09 & 0.22~$\pm$~0.15  &   0.30~$\pm$~0.16 &    0.04~$\pm$~0.14 &    0.09~$\pm$~0.11 & $-$0.10~$\pm$~0.15 & $-$0.02~$\pm$~0.15 & $-$0.07~$\pm$~0.16    \\
WZ Sgr            &     0.06~$\pm$~0.16 & 0.62~$\pm$~0.15  &   0.56~$\pm$~0.20 &    0.15~$\pm$~0.20 &    0.31~$\pm$~0.16 &    0.20~$\pm$~0.16 &    0.19~$\pm$~0.15 &    0.39~$\pm$~0.20    \\
X Pup             &  $-$0.33~$\pm$~0.14 & 0.39~$\pm$~0.15  &   0.51~$\pm$~0.15 &    0.30~$\pm$~0.20 &    0.26~$\pm$~0.13 &    0.05~$\pm$~0.15 & $-$0.16~$\pm$~0.15 &    0.04~$\pm$~0.17    \\
XX Car            &  $-$0.44~$\pm$~0.13 & 0.50~$\pm$~0.15  &   0.49~$\pm$~0.20 &    0.09~$\pm$~0.20 &    0.26~$\pm$~0.08 &    0.19~$\pm$~0.16 &    0.08~$\pm$~0.15 &    0.34~$\pm$~0.20    \\
YZ Car            &  $-$0.19~$\pm$~0.19 & 0.38~$\pm$~0.15  &   0.30~$\pm$~0.20 &    0.20~$\pm$~0.15 &    0.32~$\pm$~0.10 &    0.20~$\pm$~0.17 &    0.08~$\pm$~0.15 &    0.20~$\pm$~0.18    \\
OGLE LMC-CEP-0512 &  $-$0.12~$\pm$~0.13 &      $---$       &$-$0.60~$\pm$~0.14 & $-$0.34~$\pm$~0.16 & $-$0.15~$\pm$~0.08 & $-$0.06~$\pm$~0.13 &       $---$        & $-$0.40~$\pm$~0.17    \\
OGLE LMC-CEP-0992 &  $-$0.31~$\pm$~0.18 &      $---$       &$-$0.33~$\pm$~0.16 & $-$0.15~$\pm$~0.20 & $-$0.10~$\pm$~0.16 & $-$0.04~$\pm$~0.14 & $-$0.25~$\pm$~0.15 & $-$0.21~$\pm$~0.14    \\
\hline
&&&&&&&&\\
    ID             &         [K/H]~~~~  &  [Ca/H]~~~~        &    [Ti/H]~~~~      &  [Mn/H]~~~~        &   [Fe/H]~~~~        &     [Ni/H]~~~~      &     [Ce/H]~~~~      &     [Yb/H]~~~~        \\
\hline
AQ Car             &    0.14~$\pm$~0.15 & $-$0.18~$\pm$~0.13 & $-$0.10~$\pm$~0.15 & $-$0.22~$\pm$~0.15 &     0.00~$\pm$~0.14 &  $-$0.15~$\pm$~0.20 &  $-$0.30~$\pm$~0.15 &   0.45~$\pm$~0.15 \\
AQ Pup             &    0.18~$\pm$~0.15 & $-$0.20~$\pm$~0.12 & $-$0.03~$\pm$~0.15 &    0.03~$\pm$~0.15 &  $-$0.04~$\pm$~0.14 &  $-$0.12~$\pm$~0.17 &     0.10~$\pm$~0.15 &   0.67~$\pm$~0.15 \\
DR Vel             &    0.27~$\pm$~0.15 &    0.03~$\pm$~0.20 & $-$0.12~$\pm$~0.15 &    0.15~$\pm$~0.15 &     0.01~$\pm$~0.14 &  $-$0.09~$\pm$~0.15 &  $-$0.02~$\pm$~0.15 &   0.31~$\pm$~0.15 \\
RV Sco             &    0.07~$\pm$~0.15 & $-$0.09~$\pm$~0.14 &    0.05~$\pm$~0.15 & $-$0.03~$\pm$~0.15 &  $-$0.04~$\pm$~0.13 &     0.00~$\pm$~0.15 &  $-$0.43~$\pm$~0.15 &   0.20~$\pm$~0.15 \\
S Nor              &    0.06~$\pm$~0.15 & $-$0.10~$\pm$~0.12 & $-$0.10~$\pm$~0.15 & $-$0.08~$\pm$~0.15 &     0.03~$\pm$~0.14 &  $-$0.10~$\pm$~0.18 &  $-$0.15~$\pm$~0.15 &   0.27~$\pm$~0.15 \\
SS Cma             & $-$0.06~$\pm$~0.15 & $-$0.10~$\pm$~0.20 & $-$0.10~$\pm$~0.15 & $-$0.15~$\pm$~0.15 &  $-$0.11~$\pm$~0.19 &     0.00~$\pm$~0.15 &  $-$0.28~$\pm$~0.15 &   0.40~$\pm$~0.15 \\
S Tra              &    0.13~$\pm$~0.15 & $-$0.10~$\pm$~0.18 & $-$0.15~$\pm$~0.15 &    0.05~$\pm$~0.15 &  $-$0.02~$\pm$~0.11 &  $-$0.12~$\pm$~0.16 &  $-$0.05~$\pm$~0.15 &   0.19~$\pm$~0.15 \\
SV Vel             &    0.20~$\pm$~0.15 & $-$0.10~$\pm$~0.20 & $-$0.13~$\pm$~0.15 &    0.08~$\pm$~0.15 &  $-$0.11~$\pm$~0.14 &  $-$0.20~$\pm$~0.16 &     0.01~$\pm$~0.15 &   0.33~$\pm$~0.15 \\
SV Vul             &    0.20~$\pm$~0.15 &    0.15~$\pm$~0.16 &    0.24~$\pm$~0.15 &    0.20~$\pm$~0.15 &     0.14~$\pm$~0.14 &  $-$0.03~$\pm$~0.15 &     0.44~$\pm$~0.15 &   0.55~$\pm$~0.15 \\
T Mon              &    0.15~$\pm$~0.15 &    0.15~$\pm$~0.20 &    0.15~$\pm$~0.15 & $-$0.07~$\pm$~0.15 &     0.12~$\pm$~0.20 &  $-$0.05~$\pm$~0.15 &  $-$0.15~$\pm$~0.15 &   0.60~$\pm$~0.15 \\
U Aql              & $-$0.00~$\pm$~0.15 &    0.18~$\pm$~0.15 & $-$0.01~$\pm$~0.15 &    0.00~$\pm$~0.15 &  $-$0.05~$\pm$~0.17 &  $-$0.03~$\pm$~0.15 &     0.24~$\pm$~0.15 &   0.08~$\pm$~0.15 \\
U Car              &    0.23~$\pm$~0.15 &    0.23~$\pm$~0.20 &    0.02~$\pm$~0.15 &    0.20~$\pm$~0.15 &     0.28~$\pm$~0.08 &     0.20~$\pm$~0.15 &  $-$0.05~$\pm$~0.15 &   0.65~$\pm$~0.15 \\
U Sgr              &    0.08~$\pm$~0.15 & $-$0.05~$\pm$~0.16 & $-$0.11~$\pm$~0.15 &    0.10~$\pm$~0.15 &     0.15~$\pm$~0.16 &     0.14~$\pm$~0.15 &  $-$0.62~$\pm$~0.15 &   0.27~$\pm$~0.15 \\
V0636 Sco          &    0.16~$\pm$~0.15 & $-$0.15~$\pm$~0.14 & $-$0.20~$\pm$~0.15 &    0.10~$\pm$~0.15 &     0.03~$\pm$~0.14 &  $-$0.08~$\pm$~0.13 &  $-$0.21~$\pm$~0.15 &   0.17~$\pm$~0.15 \\
VZ Pup             &    0.14~$\pm$~0.15 &    0.05~$\pm$~0.15 &    0.00~$\pm$~0.15 & $-$0.02~$\pm$~0.15 &  $-$0.04~$\pm$~0.18 &  $-$0.03~$\pm$~0.15 &     0.28~$\pm$~0.15 &   0.75~$\pm$~0.15 \\
W Gem              &    0.00~$\pm$~0.15 & $-$0.10~$\pm$~0.18 &    0.10~$\pm$~0.15 &    0.20~$\pm$~0.15 &     0.02~$\pm$~0.17 &  $-$0.10~$\pm$~0.15 &  $-$0.54~$\pm$~0.15 &   0.24~$\pm$~0.15 \\
WX Pup             &    0.07~$\pm$~0.15 & $-$0.14~$\pm$~0.18 &    0.08~$\pm$~0.15 & $-$0.04~$\pm$~0.15 &  $-$0.09~$\pm$~0.14 &     0.00~$\pm$~0.15 &     0.16~$\pm$~0.15 &   0.70~$\pm$~0.15 \\
WZ Sgr             &    0.13~$\pm$~0.15 &    0.29~$\pm$~0.20 &    0.04~$\pm$~0.15 & $-$0.13~$\pm$~0.15 &     0.10~$\pm$~0.18 &     0.05~$\pm$~0.15 &  $-$0.11~$\pm$~0.15 &   0.53~$\pm$~0.15 \\
X Pup              &    0.30~$\pm$~0.15 &    0.08~$\pm$~0.13 &    0.01~$\pm$~0.15 &    0.05~$\pm$~0.15 &     0.06~$\pm$~0.11 &     0.00~$\pm$~0.10 &     0.31~$\pm$~0.15 &   0.55~$\pm$~0.15 \\
XX Car             &    0.01~$\pm$~0.15 & $-$0.10~$\pm$~0.15 & $-$0.04~$\pm$~0.15 & $-$0.20~$\pm$~0.15 &     0.00~$\pm$~0.17 &  $-$0.20~$\pm$~0.14 &  $-$0.13~$\pm$~0.15 &   0.59~$\pm$~0.15 \\
YZ Car             &    0.20~$\pm$~0.15 &    0.00~$\pm$~0.15 & $-$0.05~$\pm$~0.15 &    0.08~$\pm$~0.15 &     0.13~$\pm$~0.15 &     0.00~$\pm$~0.17 &  $-$0.05~$\pm$~0.15 &   0.58~$\pm$~0.15 \\
OGLE LMC-CEP-0512  &        $---$       & $-$0.21~$\pm$~0.16 &         $---$      & $-$0.30~$\pm$~0.15 &  $-$0.39~$\pm$~0.20 &           $---$     &     0.00~$\pm$~0.15 &   0.80~$\pm$~0.15 \\
OGLE LMC-CEP-0992  & $-$0.28~$\pm$~0.15 & $-$0.10~$\pm$~0.20 & $-$0.12~$\pm$~0.15 & $-$0.28~$\pm$~0.15 &  $-$0.31~$\pm$~0.15 &  $-$0.81~$\pm$~0.20 &     0.50~$\pm$~0.15 &   0.50~$\pm$~0.15 \\
\hline
\end{tabular}
\end{sidewaystable*}

\end{appendix}

\end{document}